\documentclass{aa}  

\usepackage{graphicx}
\usepackage{txfonts}
\usepackage[final]{hyperref}
\usepackage{multicol}
\usepackage{multirow}
\usepackage{placeins}

\makeatletter
\renewcommand*\aa@pageof{, page \thepage{} of \pageref*{LastPage}}
\makeatother

\begin{document}

   \title{Theoretical computations on the efficiency of acetaldehyde formation on interstellar icy grains}


   \author{Joan Enrique-Romero\inst{1,2}
          \and
          Cecilia Ceccarelli\inst{1}
          \and
          Albert Rimola\inst{2}
          \and
          Dimitrios Skouteris\inst{3}
          \and
          Nadia Balucani\inst{1,4,5}
          \and
          Piero Ugliengo\inst{6}
          }

   \institute{Univ. Grenoble Alpes, CNRS, Institut de Plan\'{e}tologie et d'Astrophysique de Grenoble (IPAG), 38000 Grenoble, France\\
              \email{juan.enrique-romero@univ-grenoble-alpes.fr}\\
              \email{cecilia.ceccarelli@univ-grenoble-alpes.fr}
         \and
             Departament de Qu\'{i}mica, Universitat Aut\`{o}noma de Barcelona, Bellaterra, 08193, Catalonia, Spain\\
             \email{albert.rimola@uab.cat}
         \and
            Master-Tech, I-06123 Perugia, Italy
         \and
            Dipartimento di Chimica, Biologia e Biotecnologie, Universit\`{a} di Perugia, Via Elce di Sotto 8, 06123 Perugia, Italy
         \and
            Osservatorio Astrofisico di Arcetri, Largo E. Fermi 5, 50125 Firenze, Italy
         \and
            Dipartimento di Chimica and Nanostructured Interfaces and Surfaces (NIS) Centre, Universit\`{a} degli Studi di Torino, via P. Giuria 7, 10125, Torino, Italy
             }

   \date{Received --; accepted --}

 
  \abstract
   {Interstellar grains are known to be important actors in the formation of interstellar molecules such as H$_2$, water, ammonia, and methanol. 
   It has been suggested that the so-called interstellar complex organic molecules (iCOMs) are also formed on the interstellar grain icy surfaces by the combination of radicals via reactions assumed to have an efficiency equal to unity.}
   {In this work, we aim to investigate the robustness or weakness of this assumption.
   In particular, we consider the case of acetaldehyde (CH$_3$CHO), one of the most abundant and commonly identified iCOMs, as a starting study case.
   In the literature, it has been postulated that acetaldehyde is formed on the icy surfaces via the combination of HCO and CH$_3$.
   Here we report new theoretical computations on the efficiency of its formation.}
   {To this end, we coupled quantum chemical calculations of the energetics and kinetics of the reaction CH$_3$ + HCO, which can lead to the formation of CH$_3$CHO or CO + CH$_4$.
   Specifically, we combined reaction kinetics computed with the Rice–Ramsperger–Kassel–Marcus (RRKM) theory (tunneling included) method with diffusion and desorption competitive channels.
   We provide the results of our computations in the format used by astrochemical models to facilitate their exploitation.}
   {Our new computations indicate that the efficiency of acetaldehyde formation on the icy surfaces is a complex function of the temperature and, more importantly, of the assumed diffusion over binding energy ratio $f$ of the CH$_3$ radical.
   If the ratio $f$ is $\geq$0.4, the efficiency is equal to unity in the range where the reaction can occur, namely between 12 and 30 K.
   However, if $f$ is smaller, the efficiency dramatically crashes: with $f$=0.3, it is at most 0.01.
   In addition, the formation of acetaldehyde is always in competition with that of CO + CH$_4$.}
   {Given the poor understanding of the diffusion over binding energy ratio $f$ and the dramatic effect it has on the formation, or not, of acetaldehyde via the combination of HCO and CH$_3$ on icy surfaces, model predictions based on the formation efficiency equal to one should to be taken with precaution.
   The latest measurements of $f$ suggest $f$=0.3 and, if confirmed for CH$_3$, this would rule out the formation of acetaldehyde on the interstellar icy surfaces.
   We recall the alternative possibility, which was recently reviewed, that acetaldehyde could be synthesized in the gas phase starting from ethanol.
   Finally, our computations show the paramount importance played by the micro-physics involved in the interstellar surface chemistry and call for extensive similar studies on different systems believed to form iCOMs on the interstellar icy surfaces.
}


   \maketitle
%
\section{Introduction}\label{sec:introduciton}

Interstellar dust grains are known to be an essential component of the interstellar medium (ISM) for a large variety of reasons. 
Among them, dust grains provide the surfaces for chemical reactions that are difficult (or impossible) to take place in the gas phase. 
An emblematic example is the formation of the most abundant molecule in the universe, H$_2$, which largely occurs on grain surfaces  \citep[e.g.,][]{hollenbach1970,Vidali2013,Wakelam2017H2}.
Other important examples are the formation of water \citep[e.g.,][]{dulieu2010,he2014,Lamberts2017,Molpeceres2019} and methanol \citep[e.g.,][]{tielens1982,Watanabe2002,rimola2014}, which are also abundant molecules predominantly synthesized on grain surfaces.
As a matter of fact, in cold regions, the refractory core of the grains, made up of silicate or carbonaceous material, are coated by icy mantles mostly formed by amorphous water ice synthesized on these surfaces \citep[e.g.,][]{Boogert2015}.

It has also been suggested that more complex molecules, the so-called interstellar complex organic molecules \citep[hereinafter iCOMs:][]{ceccarelli2017}, composed of at least six atoms and containing at least one heavy element other than C, can also be a grain-surface chemistry product \citep[e.g.,][]{Garrod2006,Herbst2009,Garrod2009,Ruaud_ER_2015,aikawa2020,BargerGarrod2020}.
One crucial step of this theory is the formation of iCOMs from the combination of two radicals when they meet on the grain icy surfaces. 
In the majority of the current astrochemical models, the reaction is assumed to proceed barrierless and without competitive channels. 

However, previous theoretical works have shown that this is not necessarily the case \citep[][]{Rimola2018,ER2019,ER2020}.
For example, theoretical calculations on the energetics showed that the formation of acetaldehyde (CH$_3$CHO) on the icy surfaces via HCO + CH$_3$ is in competition with the formation of CO + CH$_4$ via direct H-abstraction.
In addition, both reactions present barriers, caused by the orientation of the species on the ices, which are governed by the interactions created between the surface water molecules and the two radicals.
Indeed, the height of the barriers depends on the site where the reaction occurs, whether the two radicals are on a ``plain" ice surface or in a ``cavity", namely on the interactions between the radicals and the ice water molecules.

In the present work, we pursue the above theoretical studies and present new computations to evaluate the efficiency of the radical-radical combination and H-abstraction reactions as a function of the temperature of the icy surfaces with the goal to provide values that can be easily incorporated in astrochemical models.
In particular, here we focus on acetaldehyde (CH$_3$CHO), one of the most abundant and common iCOMs \citep[e.g.,][]{Blake1987,cazaux2003,vastel2014,lefloch2017,Sakai2018,Bianchi2019,Csengeri2019,Lee2019,Scibelli2020}, as a study case.

Following the works of \cite{ER2019,ER2020}, we here consider the two competing reactions that can arise from the reactivity between HCO and CH$_3$:

\begin{tabular}{cl}
    (1) & HCO + CH$_3$ ~$\to$~ CH$_3$HCO \\
    (2) & HCO + CH$_3$ ~$\to$~ CH$_4$ + CO. 
\end{tabular}

Our goal is to evaluate the efficiency of each of the two reactions occurring on the interstellar icy surfaces.
To this end, we computed the kinetics of the two reactions, using the previous energetic calculations by \cite{ER2019} as a base and the models of amorphous solid water (ASW) for the ice described in \cite{Rimola2018} and \cite{ER2019}.
In our calculations, we assume that the two radicals are in the most stable energetic configuration prior to reaction, an assumption motivated by the long surface residence timescale \citep[1--10 Myr, the molecular life timescale: e.g.][]{Chevance2020} that radicals would experience before they become mobile and react with each other. 

The article is organized as follows.
The definition of the reaction efficiency and our choices for the various assumptions entering in the computations are discussed in Sect. \ref{sec:def-probability}. 
Sect. \ref{sec:methodology} describes the adopted methodology.
The results are reported in Sect. \ref{sec:results} and we discuss the implications of our new calculations in Sect. \ref{sec:discussion}.

\section{Efficiency of radical-radical reaction products on icy surfaces}\label{sec:def-probability}

\subsection{Surface-reaction rate definition}\label{subsec:reac-rate-def}

Generally, astrochemical models solve the time-dependent equations of the species densities by computing the formation and destruction rates of each species at a given time, both for species in the gas and on the grain surfaces.
In particular, the rate $R_{ij}$ of the formation reaction from two reactant species $i$ and $j$ is expressed as $R_{ij}=k_{ij}n_in_j$, where $n_i$ and $n_j$ are the densities of species $i$ and $j$, and $k_{ij}$ is the rate constant at a given temperature. For surface reactions the latter is given by \citep{HHL1992}: 

\begin{equation}
k_{ij} =\varepsilon_{ij} \times \frac{R_{\text{diff},i} + R_{\text{diff},j}}{n_d},
    \label{eqn:rate_rx}
\end{equation}

where $\varepsilon_{ij}$ is an efficiency factor which accounts for chemical barriers, $n_d$ is the dust grain density and $R_{\text{diff},i}$ and $R_{\text{diff},j}$ are the diffusion rates for species $i$ and $j$, respectively. 
These diffusion rates are defined as $1/t_{\text{diff},k}$, where $t_{\text{diff},k}$ is the time it takes the species $k$ to scan the whole grain \citep[e.g.,][]{Garrod2006}. 
Thus, the sum $R_{\text{diff},i} + R_{\text{diff},j}$ gives the rate at which species $i$ and $j$ meet on the surface.

Regarding the efficiency factor $\varepsilon_{ij}$, different approaches exist in order to derive it. 
\citet{HHL1992} set it to either 1, in barrierless reactions, or to the tunnelling probability, if the reaction has an activation energy barrier and one of the reactants is light enough to tunnel through it. 
Later models include also the thermal probability for reaction, if there is an activation energy barrier \citep[e.g.,][]{Garrod2006}. 
However, in the presence of an activation energy barrier, reactants need to be close to each other for a certain amount of time for the reaction to occur \citep[][]{tielens1982}. 
In order to take this into account, \citet{Chang2007} redefined the efficiency taking into account the competition between diffusion and desorption of the most mobile species, as follows:
 \begin{equation}
 \varepsilon_{ij} = \frac{k_{aeb}(ij)}{k_{aeb}(ij) + k_{diff}(i) + k_{des}(i)},    
 \label{eqn:chang2007}
 \end{equation}
 \noindent where $k_{aeb}(ij)$ is the rate constant accounting for the reaction activation energy barrier, which is described by either classical thermal kinetics or quantum tunnelling;
(i.e., a frequency times a Boltzmann factor or the tunnelling probability);
$k_{diff}(i)$ is the rate constant for the diffusion 
of the most mobile species and $k_{des}(i)$ is its desorption rate constant. 
\citet{GP2011} further modified Eq. (\ref{eqn:chang2007}) 
by removing the desorption term and adding the diffusion of the other reaction partner, $j$, in the denominator.

For reactions involving radicals, $\varepsilon_{ij}$ is normally assumed equal to 1 \citep[e.g.,][]{Garrod2006}, as they are considered to react via barrierless exothermic channels.

In this work, we include diffusion and desorption rates of the two reactants, which takes into account both the \citet{Chang2007} and \citet{GP2011} recipes:
%
\begin{equation}
\varepsilon_{ij} = 
    \frac{k_{aeb}(ij)}
    {k_{aeb}(ij) + 
    k_{diff}(i) + k_{des}(i) +
    k_{diff}(j) + k_{des}(j)}.
    \label{eqn:our_rx_efficiency}
\end{equation}
%
In practice, the efficiency for the reaction is equal to unity only when the time scale for the reaction to occur ($1/k_{aeb}$) is shorter than the timescales at which reactants remain on the reaction site (the smallest between $1/k_{diff}$(i) and $1/k_{diff}$(j)).

\subsection{A novel treatment of surface radical-radical reactions rate constants}

The novelty of the present work is the estimate of the $k_{aeb}(ij)$ coefficient of radical-radical reactions via statistical kinetics calculations based on the Ramsperger-Rice-Kassel-Marcus (RRKM) microcanonical transition state  theory. 
Briefly, RRKM computations provide unimolecular rate constants, namely the rate at which a system A becomes A$'$ passing through a transition state (only once).
In our case, the system A is the ice-water molecules plus the two adsorbed radicals, namely we consider the water-cluster plus the radicals as a super-molecule isolated from its surrounding.
The system A$'$ is the product of the radical-radical reaction on the icy surface, namely the water-cluster plus either the radical-radical recombination (e.g., React. I) or the H-abstraction  (e.g., React. II) products.

It is important to note that in order to apply the RRKM theory, we implicitly assume that the intra-molecular energy redistribution of the reaction energy is faster than the reaction itself. 
This assumption is supported by recent ab initio molecular dynamics (AIMD) computations that show that a large fraction ($\geq50$\%) of the reaction energy is absorbed by the water ice in less than 1 ps (\cite{pantaleone2020, pantaleone2021}). 
We have checked a posteriori that the timescale of the reactions studied here is indeed longer that 1 ps. Finally, the specific computational details of our proposed RRKM method are reported in \S ~\ref{subsec:kinetic_calcs}.

\subsection{Desorption and diffusion energies}\label{subsec:des-hop-en}

Equation \ref{eqn:our_rx_efficiency} shows that, in addition to the probability $k_{aeb}(ij)$ for radicals $i$ and $j$ to react when they meet on a surface site, the efficiency factor $\varepsilon_{ij}$ also depends on $k^{-1}_{des}$ and on $k^{-1}_{diff}$, which are related  to the residence time of the radicals on the surface and on the diffusion timescale of the radicals on the ice, respectively.
The diffusion and desorption timescales $t_{\text{diff/des}}$ are given by the classical Eyring transition state theory (TST), in which $t_{\text{diff/des}}$ is inversely proportional to their rate constant, $t_{\text{diff/des}}$ $\propto  k^{-1}_{\text{diff/des}}$. According to TST, the general expression of the rate constant k for a unimolecular reaction (like diffusion and desorption) is:

\begin{equation}
k = \frac{k_B T}{h} \frac{Q^{\neq}}{Q_R} \exp(-\Delta V^{\neq}/k_B T),
    \label{eqn:Eyring-Q}
\end{equation}
where $\Delta V^{\neq}$ is the zero point energy-corrected energy barrier, $Q^{\neq}$ and $Q_{\text{R}}$ are the total partition functions of the transition state and the initial state (namely, the reactants), respectively, $k_B$ is Boltzmann's constant, $T$ is the surface temperature and $h$ is Planck's constant. We note that we use the classical Eyring equation because, since we are not dealing with light atoms but molecular radicals, tunneling is negligible. 

By proper manipulation of equation \ref{eqn:Eyring-Q}, the rate constant becomes expressed as a function of the free energy barrier $\Delta G^{\neq}$ (usually referred to as free energy of activation) at a given temperature:
\begin{equation}
k = \frac{k_B T}{h} \exp(-\Delta G^{\neq}/k_B T),
    \label{eqn:Eyring-G}
\end{equation}
in which $\Delta G^{\neq}$ = $\Delta H^{\neq}$ - T$\Delta S^{\neq}$ and where $\Delta H^{\neq}$ is the enthalpy of activation and $\Delta S^{\neq}$ the entropy of activation. These terms contain translational, rotational, vibrational and electronic contributions as they arise partly from the total partition functions Q. 

With the adopted quantum chemical approach, the application of the Eyring TST allows us to compute desorption-related data (e.g., desorption activation energies and desorption rate constants) for each radical through the outcome of these calculations (electronic energies, vibrational frequencies, partition functions, energy contributions, etc.). It is worth mentioning that, since the radicals are physisorbed on the ice surfaces, the energy barriers of the desorption processes coincide with the desorption energies. In the present case, we only account for the electronic and vibrational contributions to both $\Delta H^{\neq}$ and $\Delta S^{\neq}$ to arrive at the radical desorption energies as follows: 
\begin{equation}
\Delta H = \Delta E_{electronic} + \Delta ZPE + \Delta E_{vib}(T) + \Delta H_{rot} + \Delta H_{trans},
    \label{eqn:H-contributions}
\end{equation}
and
\begin{equation}
\Delta S = \Delta S_{vib} + \Delta S_{rot} + \Delta S_{trans},
    \label{eqn:S-contributions}
\end{equation}
where the terms are the energy difference between the desorbed and the adsorbed states for the total electronic energy ($\Delta E_{electronic}$), for the zero point vibrational energy corrections ($\Delta ZPE$), for the thermal vibrational energy corrections ($\Delta E_{vib}(T)$), for the vibrational entropy ($\Delta S_{vib}$), and for the rotational and translational contributions to enthalpy ($\Delta H_{rot}$ and $\Delta H_{trans}$, respectively) and entropy ($\Delta S_{rot}$ and $\Delta S_{trans}$, respectively). In this case, since we are dealing with the desorption of the radicals, the translational and rotational contributions arise from only the desorbed (free) radicals. Specific details on the calculation of some of these terms are provided in Appendix ~\ref{sec:thermostatistics}.
For the sake of simplicity, we refer to this final desorption energy as $E_{des}$.

In contrast to desorption, obtaining diffusion-related data with the present calculations is a daunting task, as it requires localizing a large number of transition states for the radical hoping between the different binding sites. Moreover, the use of a relatively small cluster model dramatically constraints the validity of these results because of its limition in terms of size and surface morphology. Therefore, to obtain a value for the diffusion energy of each radicals, which by analogy we will refer to as $E_{diff}$, we resorted to what is usually done in astrochemical modeling, that is, $E_{diff}$ is taken to be a fraction $f$ of $E_{des}$. However, deriving the value of $f$ has proven to be difficult, both theoretically and experimentally.
In the published astrochemical models, one can find a quite wide range of adopted $f$ values, from 0.3 to $\sim$0.8 \citep[e.g.,][]{HHL1992,Ruffle_Herbst2000}.
Some authors have taken a middle point by setting this ratio to 0.5 \citep[e.g.,][]{Garrod2006,Garrod2008b,GP2011,Ruaud_ER_2015,Vasyunin2017, Jensen2021}. 

In the past few years, theoretical and experimental works on the diffusion process of species on ASW surfaces have provided constraints to the $f$ value (see also the more extensive discussion in Sect. \ref{subsec:disc-astro-impli}).
In a theoretical work, \citet{Karssemeijer_2014} showed that the range for the $E_{diff}$/$E_{des}$ ratio can be narrowed down to 0.3--0.4 for molecules like CO and CO$_2$. 
\cite{Minissale2016} experimentally found that the $f$ ratio of atomic species like N and O is about 0.55, while \cite{He_Vidali_2018} showed that $f$ is 0.3--0.6, being the lower values more suitable for surface coverage lower than one mono-layer. 

Given the uncertainty on the $E_{diff}$/$E_{des}$ ratio for CH$_3$ and HCO, we carried out our calculations for three values: 0.3, 0.4 and 0.5.

\subsection{Ice model}

Regarding the amorphous solid water (ASW) model, there is still little knowledge that constrains the actual internal structure of interstellar ices. 
Observations suggest that the interstellar water ice is predominantly in the amorphous form \citep[e.g.,][]{Smith1989,Boogert2015} \citep[with some exceptions: e.g.][]{Molinari1999}.
Many laboratory studies have been carried out to characterize the possible porosity of the interstellar ices.
Typically, laboratory experiments produce porous ices of different densities by condensation of water vapor, even though they probably do not reproduce the interstellar water ice, in which water is believed to form \textit{in situ} by hydrogenation reactions of frozen O, O$_2$ and O$_3$ \citep[e.g.,][]{dulieu2010,HamaWatanabe2013,HeVidali2014,Potapov2021}.
In general, porous ices are detected in laboratory via the infrared (IR) signature of dangling OH groups, which are, however, missing in interstellar samples \citep{bar-nun1987,Keane2001}, \citep[see also the discussion in e.g.][]{HamaWatanabe2013,zamirri2018}. 
Several hypothesis have been suggested to explain the absence of the OH dangling signature \citep{Oba2009,Palumbo2006,Palumbo2010}, so that, at the end, there is consensus in the community that interstellar water ices are amorphous and porous in nature, even though many details are missing and we do not have a precise picture of the degree of porosity \citep[e.g.,][]{HamaWatanabe2013,Iskoski2014,Potapov2021}.

In order to simulate the interstellar icy surfaces, \citet{ER2019} considered a cluster of 33 water molecules.
This ice model possesses two major types of surface with respect to the binding capability: a cavity, where species are in general more strongly bonded to the surface, and an elongated side \citep{rimola2014,Rimola2018,ER2019}. 
In this work, we only report the analysis of the reaction occurring in the cavity for the following reason.
In astrochemical models, the vast majority of radical-radical reactions take place inside the bulk of the ice \citep[e.g.,][]{Garrod2006}. 
Therefore, the cavity site is a better representation of the sites where radical-radical reactions occur than that on the elongated side, which would at best describe the ice layer exposed to the gas and where just a tiny fraction of the  reactions can occur, considering that the ice is constituted by more than 100 layers \citep[e.g.,][]{taquet2012,aikawa2020}.

Therefore, in this work, we use the \citet{ER2019} ice model and methodology, but we improve the calculations for a better accuracy of the computed energetics, including dispersion, as described in detail in \S ~\ref{sec:methodology}.

\section{Methodology}\label{sec:methodology}
\subsection{Electronic structure calculations}\label{subsec:energetics_calcs}

Given the importance of inter-molecular interactions in radical-radical reactions, we recomputed the stationary points of the potential energy surfaces (PES) previously reported by \citep{ER2019} using the Grimme's D3 dispersion term including the Becke-Johnson damping (D3(BJ)) \citep{D3-grimme2010,d3bj_grimme}, this way improving the description of the dispersion forces with respect to the previous work.

All DFT calculations were performed with the {\sc Gaussian16} program package \citep{g16}. A benchmark study showed that the BHLYP hybrid density functional method is the best suited DFT method to study these reactions, with an average error of 3\%, and a maximum error of 5.0\% with respect to benchmark multi-reference CASPT2(2,2) calculations using {\sc OpenMolcas 18.09} (see Annex). We have chosen CASPT2(2,2) as the minimum level of post-HF theory, aware of the fact that a CASPT2 inclusive of full valence states would have been much better. The latter is prevented, however, by the size of our system.

Thus, stationary points were fully optimized using BHLYP \citep{bhandhlyp-becke1993,LYP88}\, combined with the standard 6-31+G(d,p) Pople basis set alongside the D3(BJ) dispersion term \citep{D3-grimme2010,d3bj_grimme}. When needed, intrinsic reaction coordinate (IRC) calculations at the optimization theory level were carried out to ensure that the transition states connect with the corresponding minima.
To balance the computational cost and chemical accuracy, reaction energetics were then refined by performing full BHLYP-D3(BJ)/6-311++G(2df,2pd) single-point energy calculations on the BHLYP-D3(BJ)/6-31+G(d,p) optimized stationary points. 
Improving chemical accuracy is a fundamental aspect when aiming at providing kinetic calculations and rate constants (including tunneling effects) \citep{AlvarezBarcia2018}, as in the present work. 
Additionally, as shown in \citep{Rimola2018,ER2019,ER2020}, DFT is a cost-effective methodology with which a correct description of biradical systems can be achieved by using the unrestricted broken (spin)-symmetry approach \citep[e.g.,][]{Neese2004}.

All optimized stationary points were characterized by the analytical calculation of the harmonic frequencies as minima and saddle points. Thermochemical corrections computed at BHLYP-D3(BJ)/6-31+G(d,p) were included to the single point BHLYP-D3(BJ)/6-311++G(2df,2pd) potential energy values using the standard rigid-rotor and harmonic oscillator formulae in order to obtain the zero-point vibrational energy (ZPE) corrections.

\subsection{Kinetic calculations}\label{subsec:kinetic_calcs}

In order to compute the rate constants for the chemical reactions between the radical pairs, we adapted our in-house kinetic code, based on the RRKM scheme for gas-phase reactions \citep{Skouteris2018}, to the surface plus adsorbed radicals case. 
First, we obtained the microcanonical rate constant $k_{aeb}(E)$ at a given energy $E$ as:

\begin{equation}
    k_{aeb}(E)=\frac{N(E)}{h\rho(E)},
    \label{eqn:micro-can-rates}
\end{equation}

where $N(E)$ is the sum of states for the active degrees of freedom in the transition state, $\rho (E)$ is the density of states for the active degrees of freedom in the reactant, and $h$ is the Planck constant. 
Since we aim to simulate a reaction taking place on a solid surface, only vibrational degrees of freedom are taken into account.
Second, the obtained rate constants were Boltzmann-averaged in order to derive the rate constants as a function of the temperature.

For the H abstraction reaction, we took into account tunneling effects adopting the Eckart scheme via the unsymmetric potential energy barrier approach.
In order to have a chemical system of reference to compare with, we applied the same method to the well studied reaction H + CO $\to$ HCO.
In this case, the initial structures of the reaction were taken from the theoretical study by \cite{rimola2014}, which were re-optimized at the present work computational level.
Here, from the optimized transition state, intrinsic reaction coordinate (IRC) calculations were run assuming a Langmuir-Hinshelwood (LH) like reaction, contrarily to the \cite{rimola2014} original computations.
All stationary points were characterized by frequency calculations, obtaining their (harmonic) vibrational modes and their zero-point energies.
More details of these computations can be found in Appendix \ref{sec:appendixHCO}.

\section{Results}\label{sec:results}

\subsection{Energetics of the reactions}

Table \ref{tab:energetics} presents the 0 K enthalpies (i.e., potential energies plus ZPE corrections) of the studied reactions.
The improvement in the dispersion correction and the refinement of the DFT energy slightly decrease the energy barriers of each one of the reactions to form acetaldehyde and CO + CH$_4$ by less than 2.5 kJ mol$^{-1}$ with respect to the values quoted by \cite{ER2019}.
Inversely, the H + CO $\to$ HCO reaction has a higher barrier, 13.5 kJ mol$^{-1}$, than that quoted by \cite{rimola2014}, 9.2 kJ mol$^{-1}$, for two reasons:
(i) Rimola et al. assumed an Eley-Rideal reaction 
(namely, the H atom comes from the gas phase and reacts with frozen CO), 
while here we have considered a LH mechanism (\S ~\ref{subsec:kinetic_calcs}), and (ii) they did not consider dispersion corrections.

\begin{table}[!btp]
\centering
\caption{Energetics and related parameters of the reactions and desorption and diffusion of the radicals. 
\textit{Top half:} Activation  ($\Delta$H$^{\ddagger}$) and reaction ($\Delta$H$^{RX}$) enthalpies (in kJ/mol) at 0 K (i.e., sum of electronic energies at BHLYP-D3(BJ)/6-311+G(2df,2pd)//BHLYP-D3(BJ)/6-31+G(d,p) and ZPE at BHLYP-D3(BJ)/6-31+G(d,p)) for each radical-radical reaction. Values for the H + CO ~$\to$~ HCO reference reaction are also shown. 
\textit{Bottom half:} Desorption energies ($E_{des}$) and desorption (T$_{des}$) and diffusion (T$_{diff}$) temperatures (in K) derived using the $E_{des}$ assuming diffusion-to-desorption energy ratios of 0.5, 0.4 and 0.3, see \S ~\ref{subsec:des-hop-en}.
\label{tab:energetics}}
\begin{tabular}{ccc}
\hline\hline
Product & $\Delta$H$^{\ddagger}$ & $\Delta$H$^{RX}$ \\ 
CH$_3$CHO     & 5.5  & -324.5 \\
CO + CH$_4$  & 7.2  & -328.9 \\
HCO           & 13.5 & -91.6 \\
\hline
Quantity [K] & CH$_3$ & HCO  \\ 
$E_{des}$   & 1715  & 3535 \\ 
T$_{des}$   & 30    & 68   \\ 
T$_{diff}$ (0.5)  & 15    & 32   \\
T$_{diff}$ (0.4)  & 12    & 25   \\
T$_{diff}$ (0.3)  & 9    & 19   \\
\hline\hline
\end{tabular}
\end{table}

\subsection{Rate constants}\label{subsec:rate-const}
\begin{table*}[!tbp]
\centering
\caption{Rate constants $k_{aeb}$ (in s$^{-1}$) and efficiency $\varepsilon$ of the two possible reactions between HCO and CH$_3$. 
For each reaction, we report the values of $\alpha$, $\beta$ and $\gamma$ of the rate constant $k_{aeb}$ and the efficiency $\varepsilon$ calculated assuming $E_{diff}$/$E_{des}$ equal to 0.5, 0.4 and 0.3 (first column).
The last three columns report the values of $k_{aeb}$ and $\varepsilon$ at 9, 20 and 30 K.}
\label{tab:coefficients}
\begin{tabular}{ccccccccc}
\hline\hline
 & Rate & Temperature & $\alpha$ & $\beta$ & $\gamma$ & 9 K & 20 K & 30 K \\ 
$E_{diff}/E_{des}$ & constant & [K] & [s$^{-1}$] & & [K] & & & \\ \hline \hline
\multicolumn{9}{c}{Reaction (1): HCO + CH$_3$ ~$\to$~ CH$_3$CHO}\\
&\multicolumn{5}{l}{$k_{aeb}$}              & 2.7$\times 10^{-21}$ & 1.9$\times 10^{-3}$ & 160.5 \\  
&                                      & 9--30 & 3.1$\times 10^{12}$& 0.70 & 663 & & & \\ 
\multirow{4}{*}{0.5}&\multicolumn{5}{l}{$\varepsilon$}                       & 1.0 & 1.0 & 1.0 \\
&                                      & 9--19  & 1.0 & 0.0    &  0.0        & & & \\
&                                      & 19--26 & 0.99 & -3.4$\times 10^{-3}$    &  0.06       & & & \\
&                                      & 26--30 & 0.98 & -0.01 &  0.28       & & & \\
\hline
\multirow{3}{*}{0.4}&\multicolumn{5}{l}{$\varepsilon$}                & 0.88 & 0.81 & 0.81 \\
&                                      & 9--13   & 0.43 & -0.21    &  0.28        & & & \\
&                                      & 13--30  & 1.0 & 0.14    &  -3.4        & & & \\
\hline
\multirow{2}{*}{0.3}&\multicolumn{5}{l}{$\varepsilon$}                & 4.9$\times 10^{-8}$ & 7.6$\times 10^{-3}$ & 0.01 \\
&                                      & 9--30  & 3.3 & 0.12 & 161.2  & & & \\
\hline \hline
\multicolumn{9}{c}{Reaction (2): HCO + CH$_3$ ~$\to$~ CH$_4$ + CO}\\
&\multicolumn{5}{l}{$k_{aeb}$}         & 6.8$\times10^{-5}$ & 0.05 & 20.1 \\  
&                                      & 9--15     & 6.1$\times10^{9}$  & 10.4 & -39.7  &  & & \\
&                                      & 15--30    & 1.7$\times10^{23}$  & 25.9  & -274  & & &  \\
\multirow{4}{*}{0.5}&\multicolumn{5}{l}{$\varepsilon$}                 & 1.0 & 1.0 & 1.0 \\\hline
&                                      & 9--24  & 1.0 & 0.0  & 0.0  & & & \\
&                                      & 24--28 & 0.83 & -0.12 & 2.79 & & &\\
&                                      & 28--30 & 0.53 & -0.44 & 11.4 & & & \\ \hline
\multirow{4}{*}{0.4}&\multicolumn{5}{l}{$\varepsilon$}                 & 1.0 & 1.0 & 0.35 \\
&                                      & 9--21  & 1.0 & 0.0  & 0.0  & & & \\
&                                      & 21--25 & 7.4$\times 10^{-11}$ & -13.8 & 278.8& & &\\
&                                      & 25--30 & 3.2$\times 10^{-6}$ & -5.8 & 53.3 & & & \\ \hline
\multirow{4}{*}{0.3}&\multicolumn{5}{l}{$\varepsilon$}                 & 1.0 & 0.02 & 1.5$\times 10^{-3}$ \\
&                                      & 9--15  & 0.96 & -0.01 & 0.02  & & & \\
&                                      & 15--19 & 2.4$\times 10^{-76}$ & -84.9 & 1205.1  & & & \\
&                                      & 19--30 & 2.4$\times 10^{7}$ & 19.9 & -660.4  & & & \\
\hline\hline
\end{tabular}
\end{table*}

\begin{figure}[!htbp]
    \centering
    \includegraphics[width=\columnwidth]{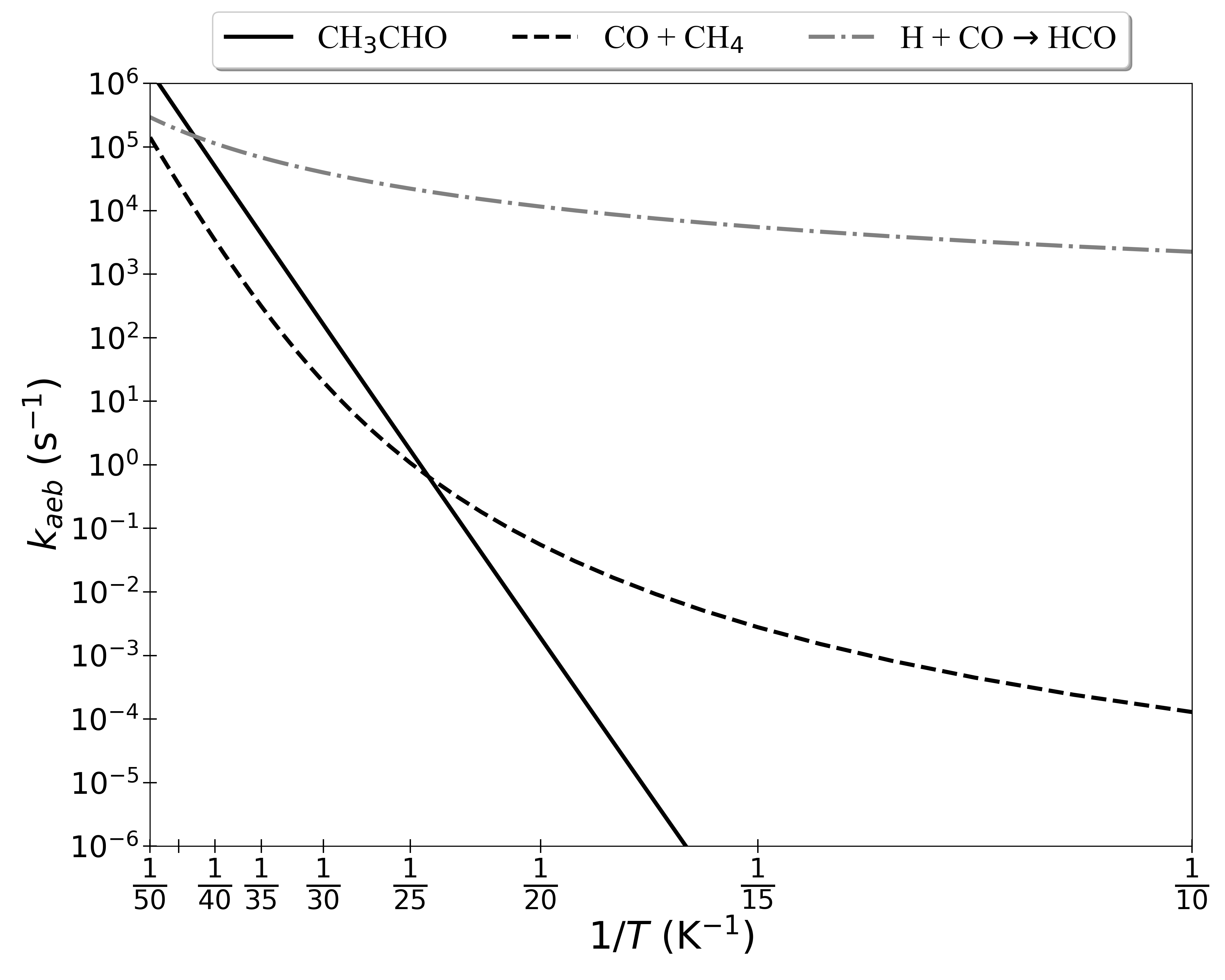}
    \caption{Arrhenius plots, namely rate constants as a function of the inverse of temperature, for the reaction CH$_3$ + HCO forming acetaldehyde (black solid line) or CO + CH$_4$ (black dashed line), and for the reaction H + CO ~$\to$~ HCO (gray dotted-dashed line), described in the main text.}
    \label{fig:unimoRatesNonDeut}
\end{figure}

Figure \ref{fig:unimoRatesNonDeut} shows the rate constants as a function of the temperature of the reactions that form CH$_3$CHO and CO + CH$_4$ from the coupling and direct H-abstraction of CH$_3$ + HCO, respectively.
The figure also reports the case of HCO formation from H + CO, for the sake of reference.

The rate constants of the reactions leading to CO + H$_2$CO and HCO take tunneling into account, which is evidenced by their deviation from linearity. 
It is also evident the strong temperature dependence of the radical-radical reactions studied, as compared to HCO formation.

The rate constants of the acetaldehyde formation are larger than those of CO + CH$_4$ formation at temperatures above $\sim$24 K. 
This is due to its lower barrier and the almost negligible quantum tunnelling contribution to HCO + CH$_3$ ~$\to$~ CO + CH$_4$ at such temperatures. 
However, as the temperature decreases the tunnelling probability takes over deviating the rate constant of CO + CH$_4$ formation from linearity, becoming faster than the formation of acetaldehyde. 
On the contrary, HCO formation has a much weaker temperature dependence and higher rate constants over the considered temperature range. 
This is the result of the dominant strong quantum tunnelling of the H atom through the reaction barrier, in agreement with the literature results \citep[e.g.,][]{Andersson2011, rimola2014}.

In order to facilitate the introduction of the new rate constants in astrochemical models, we fit the reactions rate constants with the standard formula in Eq. \ref{eqn:coefficients}. The values of $\alpha$, $\beta$ and $\gamma$ are listed in Table \ref{tab:coefficients}.
%
\begin{equation}
    k_{aeb}(T) = \alpha \left( \frac{T}{300K} \right)^\beta \exp(-\gamma/T).
    \label{eqn:coefficients}
\end{equation}

\subsection{Desorption and diffusion temperatures}\label{subsec:Tdes-Tdiff}

Table \ref{tab:energetics} reports the computed $E_{des}$ and the temperature for desorption T$_{des}$ and diffusion T$_{diff}$ derived assuming a half-life of 1 Myr. 
$E_{des}$  values are obtained at BHLYP-D3(BJ)/6-311++G(2df,2pd)//BHLYP-D3(BJ)/6-31+G(d,p) level following the procedure explained in \S ~\ref{subsec:des-hop-en}, which moreover are corrected for deformation and basis set superposition energy.
The T$_{des}$ and T$_{diff}$ values are obtained by using the standard equation for the half-life time, t$_{1/2}$=$\ln(2)$/k$_{diff/des}$(T).
These timescales provide an estimation of the characteristic temperatures for desorption and diffusion of the two radicals, CH$_3$ and HCO, involved in the formation of acetaldehyde on the icy surface.

Finally, we note that our $E_{des}$ are consistent with those computed by \cite{Ferrero2020} on a substantially larger ASW ice model.
Specifically, our $E_{des}$ in Table \ref{tab:energetics} lies in the high end of the Ferrero et al. range.
On the contrary, and as already discussed in \cite{Ferrero2020}, our $E_{des}$ are different than those reported in the astrochemical databases KIDA\footnote{\url{http://kida.astrophy.u-bordeaux.fr/}} and UMIST\footnote{\url{http://udfa.ajmarkwick.net/}}, often used by modellers.
Unfortunately, no experimental data on the CH$_3$ and HCO $E_{des}$ desorption energy exist, to our best knowledge.

\section{Discussion}\label{sec:discussion}

\subsection{\texorpdfstring{Formation of acetaldehyde versus CO + CH$_4$}{Formation of acetaldehyde versus CO + CH4}}\label{subsec:aceta-formation}

\begin{figure}[!tbp]
    \centering
    \includegraphics[width=\columnwidth]{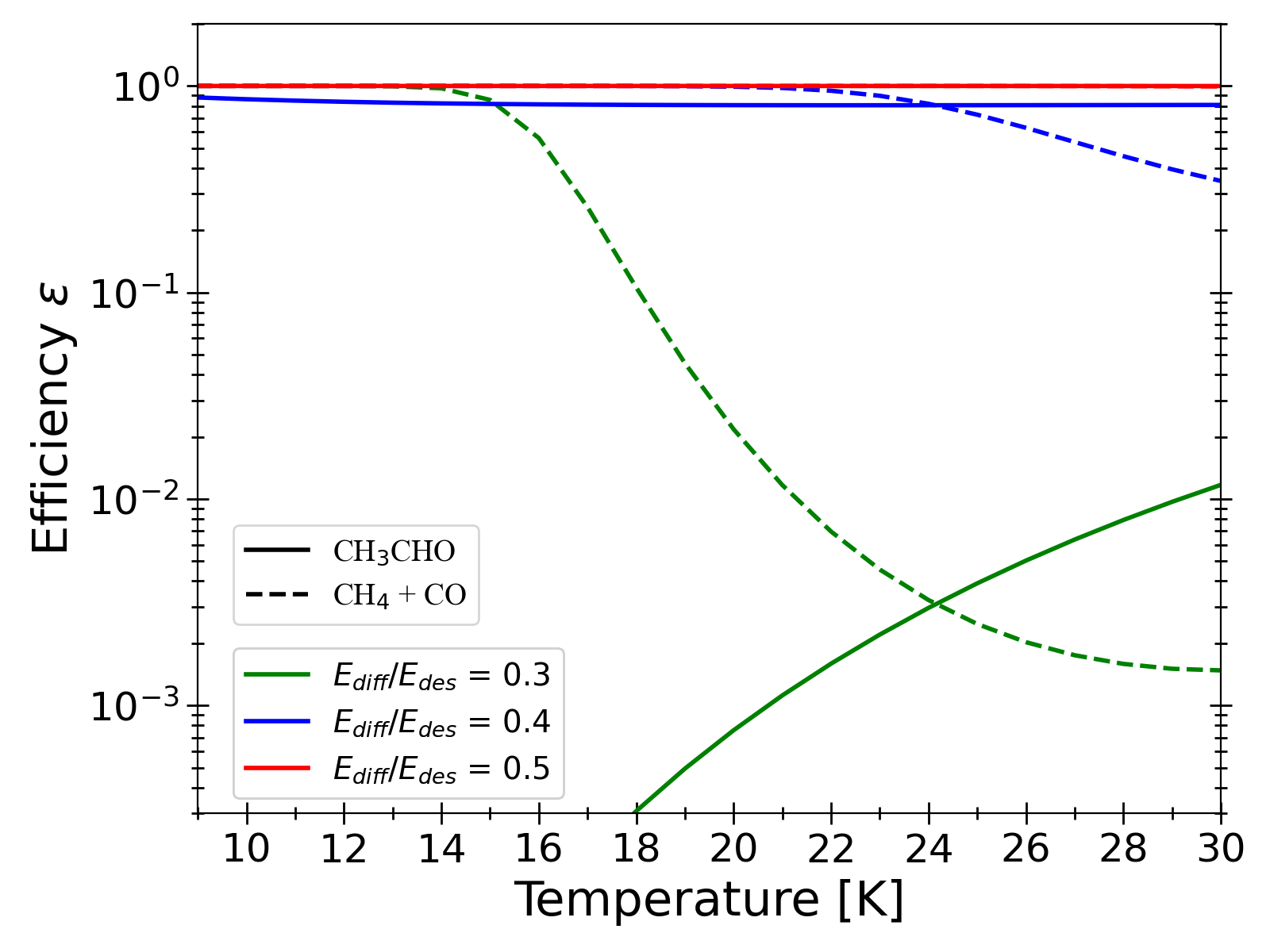}
    \caption{Reaction efficiency $\varepsilon$ (Eq. \ref{eqn:our_rx_efficiency}) of the reaction CH$_3$ + HCO leading to either CH$_3$CHO (solid lines) or CO + CH$_4$ (dashed lines) as a function of the temperature. 
    The computations were obtained adopting three different $E_{diff}$/$E_{des}$ ratios: 0.3 (green), 0.4 (blue) and 0.5 (red). 
    We note that, for $E_{diff}$/$E_{des}$=0.5 the CH$_3$CHO and CO + CH$_4$ (red) curves overlap, namely they are constant and equal to 1.
    }
    \label{fig:efficiencies_w33}
\end{figure}

We used the results of our new calculations (Sect. \ref{sec:results}) of the CH$_3$ + HCO reaction kinetics, desorption and diffusion rate constants to compute the efficiency $\varepsilon$ (Eq. \ref{eqn:our_rx_efficiency}) of the two channels leading to the formation of either CH$_3$CHO or CO + CH$_4$.
As discussed in Sect. \ref{subsec:des-hop-en}, given the uncertainty on its value, we considered three cases for the $E_{diff}$/$E_{des}$ ratio $f$: 0.3, 0.4 and 0.5.
Figure \ref{fig:efficiencies_w33} shows the resulting $\varepsilon$ as a function of the temperature and Table \ref{tab:coefficients} reports the $\alpha$, $\beta$ and $\gamma$ values obtained by fitting the $\varepsilon$ curves with Eq. (\ref{eqn:coefficients}), for the three cases of $f$.

We note that, although we computed the efficiency of the reactions in the 5--100 K range, they will only take place as long as one of the two radicals can diffuse and scan the ASW sites and both radicals stay on the reaction site, namely they do not desorb (see Table \ref{tab:energetics}).
Consequently, the upper limit to the temperature where the CH$_3$CHO and CO + CH$_4$ formation reactions take place is set by the desorption of CH$_3$, as it has a lower desorption energy than HCO (30 and 68 K, respectively, see Table \ref{tab:energetics}).
Likewise, the lower limit is also set by the CH$_3$ diffusion energy only, which is equal to 15, 12 and 9 K for $f$ equal to 0.5, 0.4 and 0.3, respectively.
In the case of $f$ equal to 0.5, HCO starts to be mobile when CH$_3$ has already sublimated, so that the efficiency of the reaction depends on CH$_3$ $E_{diff}$ only.
Conversely, for $f$ equal to 0.4 and 0.3, the temperatures at which HCO and CH$_3$ can diffuse overlap, so that both species contribute to the denominator of Eq. (\ref{eqn:our_rx_efficiency}).

\paragraph{Formation efficiency:}
For both reactions, formation of CH$_3$CHO and CO + CH$_4$, the efficiency $\varepsilon$ is about 1 in the 9--15 K range regardless of the $f$ value (between 0.3 and 0.5) with one exception, acetaldehyde formation with $f$=0.3, which starts at very low efficiency values and monotonically increases.
For $f$=0.5, either reactions have efficiencies of about unity in the whole range of temperatures (up to 30 K).
For $f$=0.4, formation of CO + CH$_4$ distances from unity at temperatures above $\sim$ 22 K, going into lower values so that at 30 K it reaches $\varepsilon\sim$0.3, while the efficiency of acetaldehyde formation stays about unity up to 30 K, where it takes a value of $\sim$0.8.
On the other hand, for $f$=0.3 things are very different. The efficiency of CO + CH$_4$ crashed at higher temperatures, reaching values of about 0.001 at 30 K, while that of acetaldehyde never goes above $\sim$ 0.01.

This is because, for relatively large $f$ values ($\geq$0.4), the most mobile radical, CH$_3$, moves slowly and the two radicals have plenty of time to react when they meet before one of them moves away: $\varepsilon$ is, therefore, close to unity.
However, when the timescale for diffusion becomes smaller than the reaction timescale (i.e., $k_{diff} \gg k_{aeb}$), CH$_3$ moves away before having the time to react and the efficiency drops below unity.
In practice, the smaller the $E_{diff}/E_{des}$ ratio, the faster CH$_3$ moves and the smaller $\varepsilon$.
However, since both $k_{diff}$ and $k_{aeb}$ have an exponential dependence on the temperature, a change in behavior occurs when the reaction activation energy $\gamma$ (Eq. (\ref{eqn:coefficients})) is similar to $E_{diff}$ and the efficiency $\varepsilon$ strongly depends on the temperature.
For the formation of acetaldehyde, $\gamma$=663 K (Table \ref{tab:coefficients}) and, therefore, the change of behavior occurs when $E_{diff}/E_{des} \sim$0.40. In these cases, the lower the temperature, the larger $k_{diff}$ with respect to $k_{aeb}$ and the smaller $\varepsilon$, as shown in Fig. \ref{fig:efficiencies_w33}.
Similar arguments hold also for the HCO + CH$_3$ $\to$ CO + CH$_4$ reaction.
The only difference is that, at low temperatures, $k_{aeb}$ deviates from the exponential law because of the kicking in of the tunneling effect that greatly increases $k_{aeb}$ (giving a negative $\gamma$ values: see Table \ref{tab:coefficients}).
Since the tunneling is more efficient for decreasing temperature, the $k_{aeb}$/$k_{diff}$ ratio decreases at increasing temperatures and, consequently, $\varepsilon$ decreases.

\begin{figure}[!tbp]
    \centering
    \includegraphics[width=\columnwidth]{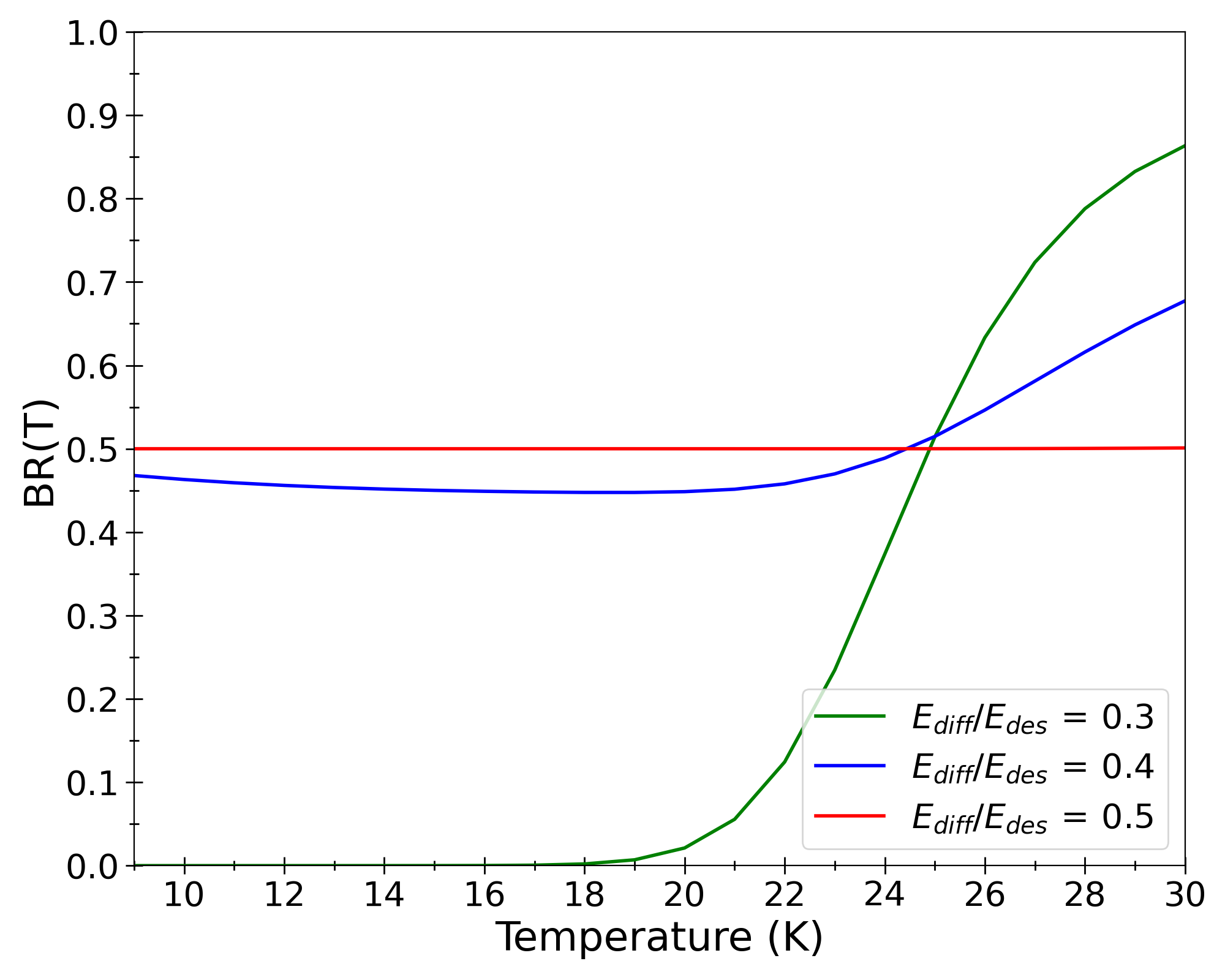}
    \caption{Branching ratio $BR(T)$ of the formation rate of the CH$_3$CHO over CO + CH$_4$ (Eq. \ref{eqn:BR}) as a function of the temperature in the range where the reactions can occur, namely below 30 K (see text), for $E_{diff}/E_{des}$ equal to 0.3 (green), 0.4 (blue) and 0.5 (red).}
    \label{fig:br}
\end{figure}

\paragraph{Branching ratio:}
Figure \ref{fig:br} shows the branching ratio $BR$ of the formation rate of CH$_3$CHO over CO + CH$_4$ as a function of the temperature, for the three $f$ values (0.5, 0.4 and 0.3). 
The $BR$ is obtained integrating Eq. (\ref{eqn:rate_rx}) from the temperature at which CH$_3$ starts to be mobile ${T_0}$, a value that depends on the assumed $f$ (see above), to the temperature $T$.
It holds:
\begin{equation}
 \text{BR}(T) = \frac{\int_{T_0}^{T} \text{d}T'\, \varepsilon_{\text{CH}_3\text{CHO}}\times (R_{\text{diff},\text{CH}_3} + R_{\text{diff},\text{HCO}} ) }
           {\int_{T_0}^{T} \text{d}T'\, (\varepsilon_{\text{CH}_3\text{CHO}} + \varepsilon_{\text{CO + CH}_4}) \times (R_{\text{diff},\text{CH}_3} + R_{\text{diff},\text{HCO}} ) }.
    \label{eqn:BR}
\end{equation}
The different effects commented above can be clearly seen in Fig. \ref{fig:br} and can be summarized as follows. 
For $f=0.5$, the branching ratio BR is constant and equal to 0.5, namely the HCO + CH$_3$ reaction leads to acetaldehyde and CO + CH$_4$ in equal quantities.
For $f=0.4$, BR lies in the range 0.4--0.5 up to 25 K and then it becomes larger, because the tunneling gain in the CO + CH$_4$ production at low temperatures vanishes.
For $f=0.3$ (and, in general, $\leq0.4$), BR is $<$0.5 at temperatures less than $\sim$25 K and rises to $\sim$0.9 at 30 K.

In other words, for $f\geq 0.4$, acetaldehyde and CO + CH$_4$ are in approximately equal competition in the range of temperatures where the HCO + CH$_3$ reaction can occur.
However, for $f< 0.4$, acetaldehyde is a very minor product for temperatures lower than about 25 K and flips to be a major product above it.

\subsection{The experimental point of view}

Experiments studying the formation of acetaldehyde from radical-radical coupling date back to the 1990s \citep{HM1997}.
They are mainly based on energetic (UV or particles) irradiation of different H$_2$O, CO, CH$_3$OH and CH$_4$ ice mixtures \cite[e.g.][]{Bennett2005,Oberg2010,MartinDomenech2020}. In relation to experimental acetaldehyde formation on grain surfaces, \cite{Bennett2005}, after irradiation of a CO:CH$_4$ ice mixture, detected acetaldehyde and predicted that the orientation of the CH$_3$ and HCO radicals are crucial in the efficiency of the reaction. On the other hand, \cite{MartinDomenech2020} conducted laboratory experiments on the formation of acetaldehyde via the CH$_3$ + HCO reaction, concluding that this channel is not efficient enough to reproduce the astronomical observations. 

As discussed by various authors, although laboratory experiments are primordial in suggesting possible mechanisms operating in the ISM and, specifically, the possible formation routes of molecules on the interstellar grain surfaces, they cannot provide the exact ISM conditions or a detailed description of the mechanisms at the atomic level. Despite this, the improvement of radical detection methods, such as the electron paramagnetic resonance (EPR) technique, will help to clarify the role of radicals generated in interstellar ice analogs \citep[e.g.,][]{Zhitnikov2002}. 
In this respect, therefore, theoretical computations as those reported in this work constitute a complementary, if not unique, tool to understand the interstellar surface chemistry.

\subsection{Astrophysical implications}\label{subsec:disc-astro-impli}

In astrochemical models, it is generally assumed that reactions between radicals on the surface of interstellar ices  are barrierless and, consequently, that their efficiency is equal to 1 (Sect. \ref{sec:def-probability}). 
In addition, it is also often assumed that there are no competition channels to the production of iCOMs.
At variance with these simple assumptions, our new calculations presented in Sect. \ref{subsec:aceta-formation}, indicate that, at low ($\leq 15$ K) temperatures, the efficiency of the acetaldehyde formation is close to unity, for a $E_{diff}/E_{des}$ ratio $f \geq$0.40.
However, there is a competing channel leading to CO + CH$_4$, for which the efficiency is also equal to 1, so that, at low temperatures and for $f \geq$0.40 the two channels are equally probable.
The acetaldehyde formation efficiency remains close to unity in the temperature range where the reaction can occur, namely at $\leq30$ K, for $f \geq$0.40.
However, the situation drastically changes for $f <$0.40.
Specifically, for $f=0.3$, the efficiency of acetaldehyde formation crashes to very low values and increases with temperature to a maximum of 0.01 at 30 K.
Similarly, the formation of CO + CH$_4$ drops to $1.5\times 10^{-3}$ at 30 K.

Therefore, two major messages come out from our calculations: (1)
the efficiency of the formation of acetaldehyde from the HCO + CH$_3$ reaction on icy surfaces is a complex function of the temperature and of the CH$_3$ diffusion energy $E_{diff}$ (Fig. \ref{fig:efficiencies_w33}) and (2) the acetaldehyde formation receives competition with CO + CH$_4$ formation, which cannot be neglected and whose efficiency is also a complex function of temperature and $E_{diff}$ (Fig. \ref{fig:br}).

While the dependence on the temperature and the importance of the competition of other products were already recognized \citep{ER2019}, the paramount importance of the diffusion energy $E_{diff}$ in the radical-radical reactions efficiency was not appreciated, at least not at the extent indicated by this study (because of the assumption of astrochemical models that the efficiency of the radical-radical reactions on grains is 1).
\cite{Penteado2017}, for example, carried out an extensive study of the surface chemistry on the binding energies (namely, our $E_{des}$) showing how critical they are.
Our new study suggests that $E_{diff}$ is as much, or even more, crucial in the reactions involving two radicals on ASW.

What makes the situation actually critical is that, while studies of the binding energy of radicals can and have been estimated in experimental and theoretical works \citep[e.g., see the recent works by][]{Penteado2017,Ferrero2020}, evaluating the diffusion energy of multi-atomic radicals on cold icy surfaces has proven to be extremely complicated and, to the best of our knowledge, no  experimental or theoretical studies exist in the literature \citep[see also e.g.,][]{Cuppen2017,Potapov2021}.
Indeed, as mentioned above, obtaining $E_{diff}$ experimentally is hitherto hampered by technical limitations on the instrumentation used to detect the radicals (i.e., EPR measurements). In relation to theoretical investigations, this lacking in bibliography is due to the convergence of methodological difficulties that make the study of diffusion with computational simulations intrinsically complex (but also compelling). Diffusion can currently be studied by means of molecular dynamics (MD) or kinetic Monte Carlo (kMC) simulations. With the first, to obtain a sufficient representativeness of the species diffusion, long simulation time-scales are mandatory. This in practice means to adopt classical force fields, in which the electronic structure of the systems is missing. However, radicals are open-shell species (with at least one unpaired electron) and accordingly electrons have to be accounted for. Thus MD simulations should be grounded within the quantum mechanics realm, which are much more expensive than the classical ones, making the MD simulations unfeasible. The alternative would be the adoption of kMC simulations. However, these simulations require building a complete network of the site-to-site radical hopping, in which for each hopping the corresponding rate constant has to be known a priori. This actually means to localize for each hopping the corresponding transition state structure (at a quantum chemical level), in which by using a realistic ASW model (i.e., large, amorphous and accordingly plenty of binding sites) makes the problem unpractical.

Usually, astrochemical models assume that the radical $E_{diff}$ is a fraction $f$ of $E_{des}$ and the value $f$ is derived from computations and experiments on species such as CO, CO$_2$, H$_2$O, CH$_4$ and NH$_3$ \citep[e.g.,][]{Mispelaer2013,Karssemeijer_2014,Lauck2015,Ghesquiere2015,He2017,Cooke2018,He_Vidali_2018,Mate2020,Kouchi2020}.
These studies give a value for $f$ between 0.3 and 0.6, as mentioned in Sect. \ref{subsec:des-hop-en}.
However, rigorously speaking, the experiments do not necessarily measure the same diffusion processes as in interstellar conditions, for at least the reasons of the surface coverage \citep{He_Vidali_2018} and its dependence of the nature of the ice, specifically its degree of porosity \citep{Mate2020}, which is poorly known in the case of interstellar ices.
As a matter of fact, using experimental and a theoretical Monte Carlo code, \cite{Mate2020} found that ``the microscopic diffusion is many times faster than the macroscopic diffusion measured experimentally".
Most recently, \cite{Kouchi2020} obtained a direct measurement of the diffusion energy of CO and CO$_2$ on ASW, using the transmission electron microscopy (TEM) technique, which allows a direct measurement of the surface diffusion coefficients (against the often used technique of IR spectroscopy, which only indirectly estimates the diffusion energy).
Kouchi and coworkers found an $f$ ratio equal to 0.3.

We have seen that, in the case of acetaldehyde, this uncertainty on $f$ has a dramatic effect.
If $f$ is $>$0.4, the efficiency of acetaldehyde formation is equal to 1 and it is about equal to that of the CO + CH$_4$ formation.
On the contrary, if $f$ is equal to 0.3, then the efficiency of acetaldehyde formation (and CO + CH$_4$) crashes, to a maximum value of 0.02.
The most recent measurements by \cite{Kouchi2020} point out the latter case as the most probable.
If the value $f$=0.3 is confirmed, then acetaldehyde is unlikely to be formed on the interstellar icy grain surfaces.

One could be tempted to use the astronomical observations against the astrochemical model predictions to add constraints to the $f$ value in (real) interstellar ices.
Of course, given the large number of parameters associated with the astrochemical models it could be a dangerous exercise.
Nonetheless, we can analyze two cases, as illustrative examples.
\citet{BargerGarrod2020} compared the predictions of their model, where the formation of acetaldehyde is dominated by the reaction CH$_3$ + HCO assumed to have $\varepsilon$=1, with the observations toward various hot cores and found that in two of them, NGC 7538 IRS 1 and W3(H$_2$O), their model overproduces the acetaldehyde column densities by more than a factor $10^3$  with respect to the observed ones.
If $f$ is equal to 0.3, introducing our new values for $\varepsilon$ could possibly cure this mismatch.
On the contrary, \citet{Jorgensen2016} found a good agreement between the observed abundances of acetaldehyde in IRAS16293B and SgrB2(N) and those predicted by the \citet{Garrod2013} model.
In this case, the agreement would point to $f \geq 0.4$.
In other words, our new computations might solve the mismatch observed toward NGC 7538 IRS 1 and W3(H$_2$O) if $f=0.3$, but they would create a mismatch on the observations toward IRAS16293B and SgrB2(N), or viceversa.
Alternatively, it is possible that $f$  varies in different sources, belonging to different environments.
For example, one could think that sources in cold quiescent regions have ices different, more or less porous, from those in warm and chaotic ones.
The two examples discussed above, unfortunately, do not lead to a coherent behavior, as, for example, IRAS16293B and SgrB2(N) could not belong to more different environments.

In conclusion, the acetaldehyde formation by radical-radical recombination on the ices is such a strong function of the diffusion energy, likely linked to the nature of the ice, that a little variation of the $E_{diff}/E_{des}$ (by 0.1) value can shift the efficiency from 1 to less than 0.01.
The most recent estimates of $E_{diff}/E_{des}$ suggest a value of 0.3 \citep{Kouchi2020}, which would make the formation of acetaldehyde on the grain surfaces unlikely.
Anyway, the important message here is that astrochemical model predictions should be taken with a certain precaution.
On the contrary, our new computations clearly show the huge importance of better knowing the microprocesses involved in the radical-radical chemistry on the icy interstellar grains and the urgent need of extensive studies, similar to the one presented here, on different systems believed to form iCOMs on the interstellar icy surfaces.

For the sake of completeness, the formation of acetaldehyde via radical-radical reactions on surfaces with lower binding energies such as solid CO could have a higher efficiency, due to the low radical--surface interactions \citep{Lamberts2019}.
Finally, it is worth reminding that acetaldehyde can alternatively be synthesized in the gas-phase \citep[e.g.,][]{Charnley2004,vastel2014,DeSimone2020_acet}. 
Recently, \cite{vazart2020} reviewed the gas-phase routes leading to acetaldehyde and found that, very likely, the dominant one is that starting from ethanol, the so-called ethanol tree \citep{Skouteris2018}.
In particular, the ethanol tree route reproduces quite well both the acetaldehyde and glycolaldehyde abundances in the sources where ethanol was also observed, including IRAS16293B \citep{vazart2020}.

\section{Conclusions}

In this work, we report new computations on the energetics and kinetics of the reaction HCO + CH$_3$, which can lead to the formation of either acetaldehyde or CH$_4$ + CO.
Specifically, we compute the rate constants of both reactions as a function of temperature as well as the efficiency of the formation of acetaldehyde and CH$_4$ + CO, respectively, combining reaction kinetics at RRKM (tunneling included) with diffusion and desorption competitive channels.
We provide analytical formulae so that the computed rate constants and efficiency can be easily introduced in astrochemical models.

The main conclusions of our study are the following.

1- The HCO + CH$_3$ reaction can only occur when the surface temperature is lower than 30 K, because CH$_3$ desorbs at larger temperatures.

2- Our computations suggest that acetaldehyde is not the dominant product for the reaction HCO + CH$_3$. The efficiency $\varepsilon$ of its formation strongly depends on the $E_{diff}/E_{des}$ ratio, providing dramatic variations between 0.3-0.5 values, the most usually used values in astrochemical models.

3- At low ($\leq 15$ K) temperatures, $\varepsilon$ is close to unity for both the formation of acetaldehyde and its competing CO + CH$_4$ channel for $f\geq$0.4, while only the efficiency of CO + CH$_4$ is unity at these temperatures for $f=0.3$ thanks to quantum tunnelling.
The efficiency of acetaldehyde formation remains unity in the range of temperatures where the reaction can occur ($\leq$30 K) for $E_{diff}/E_{des} \geq$0.40.
For lower $E_{diff}/E_{des}$ ratios, $\varepsilon$ becomes $\ll$1 and increases with increasing temperature: in the case of $E_{diff}/E_{des}$=0.3, it reaches a maximum of $\sim$0.01 at 30 K. Conversely, the efficiency of the formation of CO + CH$_4$ increases with decreasing temperature because of the tunneling. 

4- These variant $\varepsilon$ values as a function of $E_{diff}/E_{des}$ go against the assumption made in many astrochemical models, in which $\varepsilon$ is equal to 1. This might have a substantial impact on the acetaldehyde abundance predicted by these models, which may overestimate it by a few orders of magnitude.

5- We discussed the example of IRAS16293B and suggested that, in this object, acetaldehyde is likely synthesized by a gas-phase reaction route that starts from ethanol.

Finally, this new study calls for specific similar computations on the radical-radical reactions assumed to form iCOMs in astrochemical models as assuming that they have efficiency $\varepsilon$ equal to 1 and are the only reaction product could be highly misleading.


\section*{Acknowledgements}
This project has received funding within the European Union’s Horizon 2020 research and innovation programme from the European Research Council (ERC) for the projects ``The Dawn of Organic Chemistry” (DOC), grant agreement No 741002 and ``Quantum Chemistry on Interstellar Grains” (QUANTUMGRAIN), grant agreement No 865657, and from the Marie Sklodowska-Curie for the project ``Astro-Chemical Origins” (ACO), grant agreement No 811312.  
AR is indebted to ``Ram{\'o}n y Cajal" program. 
MINECO (project CTQ2017-89132-P) and DIUE (project 2017SGR1323) are acknowledged. 
Finally, we thank Prof. Gretobape for fruitful and stimulating discussions.

Most of the quantum chemistry calculations presented in this paper were performed using the GRICAD infrastructure (https://gricad.univ-grenoble-alpes.fr), which is partly supported by the Equip@Meso project (reference ANR-10-EQPX-29-01) of the programme Investissements d'Avenir supervized by the Agence Nationale pour la Recherche. Additionally this work was granted access to the HPC resources of IDRIS under the allocation 2019-A0060810797 attributed by GENCI (Grand Equipement National de Calcul Intensif).
We thank prof. Gretobape for stimulating dicsussions, and J. Perrero for useful contributions.


\bibliography{main}{}
\bibliographystyle{aa}

\twocolumn
\begin{appendix}
\section{Benchmark study}

The quality of BHLYP-D3(BJ) as an accurate, cost-effective method for the reactions studied in this work is shown in this section.

We have taken five hybrid DFT dispersion-corrected methods: MPWB1K-D3(BJ), M062X-D3, PW6B95-D3(BJ), wB97X-D3 and  BHandHLYP-D3(BJ), recommended in \cite{2017_Goerigk} for their good overall performance.

We have then compared their performance with respect to CASPT2 by studying reactions {\rm I} and {\rm II} on 2 water molecules. We proceeded in two steps: (i) we performed geometry optimized and run frequency calculations at BHLYP-D3(BJ)/6-311++G(d,p) for each reaction channel, finding and checking each stationary point (i.e., reactants, transition state and products); (ii) we then run single point calculations on top of these geometries at each DFT method combined with the 6-311++G(2df,2pd) basis set, and CASPT2/aug-cc-PVTZ for reference. The unrestricted broken symmetry scheme was adopted for all DFT calculations, and the CASPT2 guess wave function was generated using a CASSCF(2,2) calculation where the active space is formed by the two unpaired electrons and their molecular orbitals, starting from a triplet Hartree-Fock wave-function.

BHLYP-D3(BJ) gives the best overall performance with and average unsigned error of 3.0\%, and a maximum of 5.0\% with respect to CASPT2/aug-cc-PVTZ. The rest of DFT methods have average errors between 10 and 80\% (Table \ref{tab:benchmark}). The raw energy values are shown in Table \ref{tab:benchmark_raw}.

\begin{table}[!htb]
\caption{DFT method benchmark results. \% Unsigned Error from CASPT2/aug-cc-PVTZ//BHLYP-D3(BJ)/6-311++G(d,p). Methods are: (1) BHLYP-D3(BJ), (2) MPWB1K-D3(BJ), (3) M062X-D3, (4) PW6B95-D3(BJ) and (5) wB97x-D3.}
\label{tab:benchmark}
\centering
\begin{tabular}{|c|c|c|c|c|c|} 
\hline
System / Method         & 1 & 2 & 3 & 4 & 5                              \\ 
\hline
TS dHa on W2    & 5.0              & 18.5          & 59.1     & 15.8          & 46.0\\ 
\hline
HCO + CH4 on W2 & 3.5              & 2.0           & 3.9      & 1.3           & 214.3\\ 
\hline
TS Rc on W2     & 2.2              & 3.9           & 250.4    & 17.9          & 18.0\\ 
\hline
CH3CHO on W2    & 1.1              & 8.7           & 6.5      & 6.1           & 6.5\\ 
\hline
Average         & 3.0              & 8.3           & 80.0     & 10.3          & 71.2\\
\hline
\end{tabular}
\end{table}

\begin{table}[!htb]
\caption{Energetics (kJ mol$^{-1}$) of the stationary points optimized at BHLYP-D3(BJ)/6-311++G(d,p), together with the frequencies of the transition states (cm$^{-1}$).}
\label{tab:benchmark_raw_opt}
\centering
\begin{tabular}{|c|c|c|}
\hline
          & \multicolumn{2}{c|}{OPT:   BHLYP-D3(BJ)/6-311++G(d,p)} \\ \hline
          & E                            & iv                      \\ \hline
React. I  & -306.5101581                 & ~                       \\ \hline
TS I      & -306.5101000                 & -26.1495                \\ \hline
Prod. I   & -306.6472250                 & ~                       \\ \hline
React. II & -306.5106910                 & ~                       \\ \hline
TS II     & -306.5098973                 & -32.8207                \\ \hline
Prod. II  & -306.6377604                 & ~                       \\ \hline
\end{tabular}%
\end{table}

\begin{table}[!htbp]
\caption{Raw data from the DFT method benchmark. Energies in kJ mol$^{-1}$.}
\label{tab:benchmark_raw}
\centering
\resizebox{\columnwidth}{!}{%
\begin{tabular}{|c|c|c|c|c|c|}
\hline
          & \multicolumn{5}{c|}{DFT/6-311++G(2df,2pd)}                                 \\ \hline
          & BHLYP-D3(BJ) & MPWB1K-D3(BJ) & M062X-D3     & PW6B95-D3(BJ) & wB97X-D3     \\ \hline
React. I  & -306.5278673 & -306.5199277  & -306.5443283 & -307.0275371  & -306.6077421 \\ \hline
TS I      & -306.5274906 & -306.5195273  & -306.5429782 & -307.0270829  & -306.6074261 \\ \hline
Prod. I   & -306.6651096 & -306.6673671  & -306.6888795 & -307.1714705  & -306.7523061 \\ \hline
React. II & -306.5278665 & -306.5199185  & -306.5443285 & -307.0275286  & -306.6077403 \\ \hline
TS II     & -306.5273893 & -306.5195091  & -306.5435295 & -307.027106   & -306.6074694 \\ \hline
Prod. II  & -306.6561571 & -306.6555907  & -306.6825872 & -307.1623168  & -307.0257242 \\ \hline
\end{tabular}%
}
\newline
\vspace*{0.2cm}
\newline
\resizebox{0.7\columnwidth}{!}{%
\begin{tabular}{|c|c|c|}
\hline
          & \multicolumn{2}{c|}{Multi-reference methods} \\ \hline
          & CASSCF(2,2)/aug-cc-PVTZ & CASPT2/aug-cc-PVTZ \\ \hline
React. I  & -304.99849              & -306.092193        \\ \hline
TS I      & -304.99793              & -306.0918077       \\ \hline
Prod. I   & -305.14355              & -306.2278841       \\ \hline
React. II & -304.99849              & -306.0921923       \\ \hline
TS II     & -304.9976               & -306.0916902       \\ \hline
Prod. II  & -305.14411              & -306.2251986       \\ \hline
\end{tabular}%
}
\end{table}

\begin{figure}[!htbp]
    \centering
    \includegraphics[width=\columnwidth]{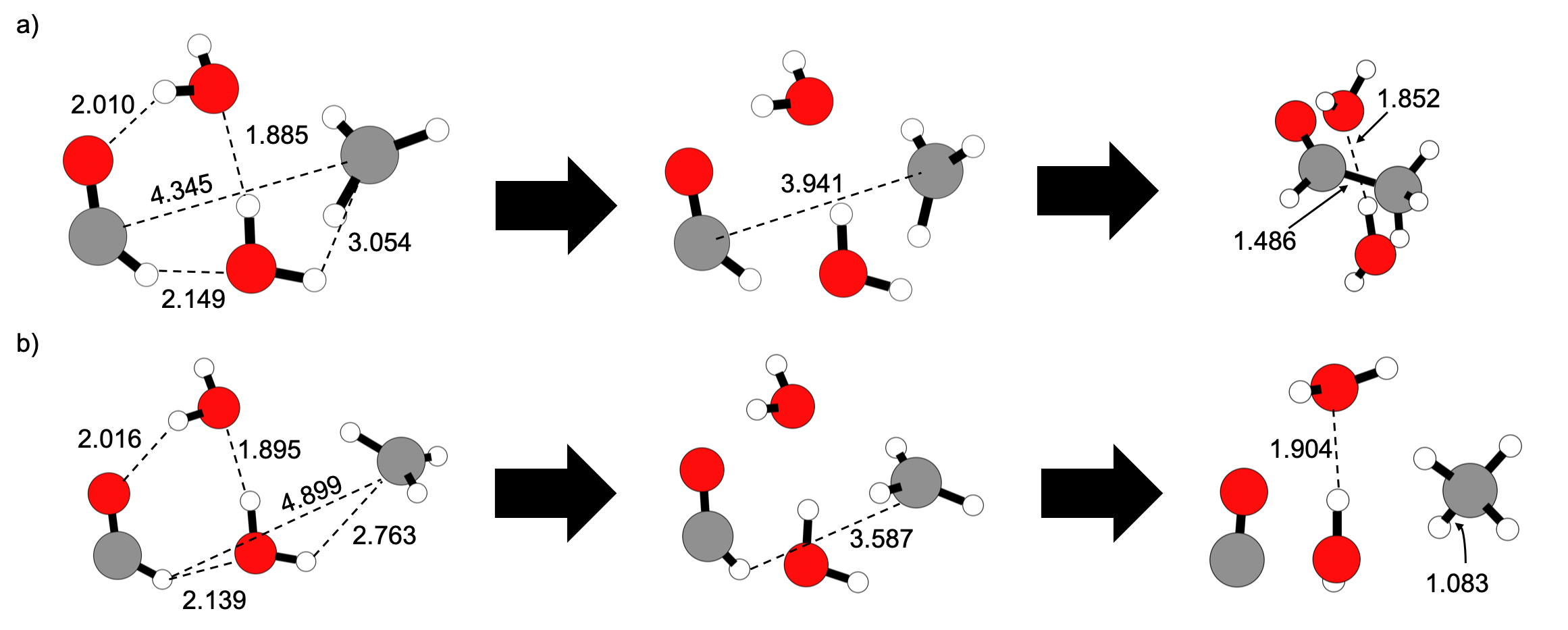}
    \caption{Geometries of the stationary points (reactants, transition state and products) of reactions a) {\rm I} and b) {\rm II} on top of two water molecules optimized at BHLYP-D3(BJ)/6-311++G(d,p) level. Distances in \r{A}.}
    \label{fig:supp:benchmark}
\end{figure}

\FloatBarrier
\section{Calculation of desorption and diffusion rate constants}\label{sec:thermostatistics}

In order to calculate the rates of diffusion and desorption we have used Eyring's equation (Eq. \ref{eqn:Eyring-G}), where entropy and thermal corrections to the enthalpy are accounted for since energy term in the exponential should be the Gibbs free energy ($G = H - TS$).
In the astrochemistry community, desorption and diffusion rates consist of two parts, first the attempt frequency (i.e., the pre-exponential factor) and then the exponential with the binding energies. 
We define the attempt frequency within the Eyring equation as

\begin{equation}
    \nu = \frac{k_BT}{h}\times \exp(\Delta S/k_B),
    \label{eqn:attempt_frqs}
\end{equation}

\noindent where $h$, $k_B$ are the Planck and Boltzmann constants, $T$ the temperature and $S$ the entropy. For the ice model and the complex surface+radical the entropy contributions are $S=S_{vib}$, where the vibrational counterpart is (Eq. \ref{eqn:supp:vibr_S}):

\begin{equation}
    S_{vib} = k_B \sum_i \left[ 
    \frac{\theta_{v,i}/T}{e^{\theta_{v,i}/T}-1} - \ln \left(
    1-e^{-\theta_{v,i}/T}
    \right)
    \right]
    \label{eqn:supp:vibr_S}
\end{equation}

\noindent where $\theta_{v,i} = hc\overline{\nu}_i/k_B$ with $c$ the speed of light and $\overline{\nu}_i$ the $i$th vibrational mode frequency in cm$^{-1}$.

For the free radicals, additionally, we also take into account the rotational and translational contributions: $S=S_{vib}+S_{rot}+S_{trans}$, where the rotational and translational counterparts are given by eqs. \ref{eqn:supp:rot_S} and \ref{eqn:supp:trans_S} respectively:

\begin{equation}
    S_{rot} = k_B \left[ 
    \ln \left( \frac{\pi^{1/2}}{\sigma_{rot}}
    \frac{T^{3/2}}{(\theta_{r,x}\theta_{r,y}\theta_{r,z})^{1/2}}
    \right)
    +\frac{3}{2}
    \right],
    \label{eqn:supp:rot_S}
\end{equation}

\begin{equation}
    S_{trans} = k_B \left( 
    \ln \left[ \left(\frac{2\pi mk_B}{h^2}\right)^{3/2}
    \frac{k_BT}{P}\right] + \frac{5}{2}
    \right),
    \label{eqn:supp:trans_S}
\end{equation}

\noindent where $\sigma_{rot}$ is the rotational symmetry number (6 for CH$_3$ and 1 for HCO), $\theta_{r,i}=B_ih/k_B$ with $B_i$ the $i$th axis rotational constant (in s$^{-1}$), $m$ is the mass of the radical and $P$ the gas pressure, which was calculated assuming a density of 10$^{4}$ cm$^{-3}$.

The thermal corrections follow: $H = E_{DFT} + ZPE + E_{vib}(T) + k_BT$ for the surface and the surface + radical complex. Here $E_{DFT}$ is the energy obtained from our DFT calculations, $ZPE$ is the zero-point energy and $E_{vib}(T)$ is the thermal vibrational energy calculated with Eq. \ref{eqn:supp:vibr_energy}. In analogy to to the entropy, the rotational and translational contributions are also included for the free radicals: $H = E_{DFT} + ZPE + E_{vib}(T) +  k_BT + H_{rot} + H_{trans}$ where $H_{rot}=H_{trans}=\frac{3}{2}k_BT$.

\begin{equation}
E_{vib}(T) = k_B \sum_i \frac{ \theta_{v,i} }{ e^{\theta_{v,i}/T}-1}
    \label{eqn:supp:vibr_energy}
\end{equation}

The magnitude of these contributions is shown in Table \ref{tab:supp:corrections}, where it can be seen that the most important contribution to the enthalpy comes from the vibrational modes, while for entropy all contributions are rather small:

\begin{table}[!htb]
\centering
\caption{Corrections incorporated in the calculation of $\Delta G$ of binding. Quantities calculated at 20 K and 30 K. Energy and entropy units are kJ mol$-1$ and kJ mol$-1$ K$^{-1}$. See that Boltzmann's constant is about 0.0083 kJ mol$^{-1}$, therefore the $k_BT$ and 3$k_BT$/2 terms are also very small.}
\label{tab:supp:corrections}
\resizebox{\columnwidth}{!}{%
\begin{tabular}{|c|c|c|c|c|c|c|c|c|}
\hline
       & \multicolumn{4}{c|}{T = 20 K}       & \multicolumn{4}{c|}{T = 30 K}       \\ \hline
       &    $\Delta E_{\text{vib}}$       & $\Delta S_{\text{vib}}$       &
            $S^{\text{rad}}_{\text{rot}}$ & $S^{\text{rad}}_{\text{trans}}$ & 
            $\Delta E_{\text{vib}}$       & $\Delta S_{\text{vib}}$       &
            $S^{\text{rad}}_{\text{rot}}$ & $S^{\text{rad}}_{\text{trans}}$ \\ \hline
CH$_3$ &    0.124                         & -0.007                        &
            0.010                         & 0.005                         &
            0.293                         & -0.003                        &
            0.015                         & 0.011                         \\ \hline
HCO    &    0.077                         & -0.012                        &
            0.034                         & 0.011                         &
            0.228                         & -0.009                        &
            0.039                         & 0.016                         \\ \hline
\end{tabular}%
}
\end{table}

As it was explained in the introduction, we have adopted the assumption that the barrier for diffusion can be expressed as a fraction of that of desorption, therefore in this model we multiply the $\Delta G$ of desorption times these fractions (assumed to be 0.3, 0.4 and 0.5).
With this, we get different attempt frequencies for diffusion and desorption (see Table \ref{tab:supp:att_frqs}), all around $10^{8}$--$10^{11}$ s$^{-1}$, smaller than the normally used approach of the harmonic oscillator\footnote{$\nu=\sqrt{2N_sE_{bind}/\pi^2 m}$, with $N_S$ the density of sites, $\sim$10$^{15}$ and $m$ the mass of the particle.} in astrochemical models (e.g., \cite{TielensHollenbach1987book,HHL1992}) due to the use of Eyring's relation and the inclusion of entropy.

\begin{table}[!htb]
\centering
\caption{Attempt frequencies for desorption and the different cases of diffusion considered in this work. Units in s$^{-1}$. See that at 20 K $k_BT/h\sim 4.2\times 10^{11}$, so the effect of entropy at such low temperatures is very small.}
\label{tab:supp:att_frqs}
\resizebox{\columnwidth}{!}{%
\begin{tabular}{|c|c|c|c|c|c|}
\hline
\multirow{2}{*}{} & \multicolumn{4}{c|}{Eyring equation (computed at 20 K)} & Harmonic \\ \cline{1-5}
                  & Desorption & \multicolumn{3}{c|}{Diffusion}   &    oscillator     \\ \hline
Diff-to-Des   & -- & 0.5 & 0.4 & 0.3 & -- \\ \hline
$\nu$(CH$_3$) & $4.7\times 10^{8}$  & $1.1\times 10^{11}$ & 
                 $1.4\times 10^{11}$ & $1.9\times 10^{11}$ & $1.5\times 10^{12}$ \\ \hline
$\nu$(HCO)    & $2.9\times 10^{10}$ & $1.4\times 10^{10}$ & 
                 $2.8\times 10^{10}$ & $5.4\times 10^{10}$ & $1.5\times 10^{12}$\\ \hline
\end{tabular}%
}
\end{table}


\FloatBarrier
\section{Rate constant comparison}

\begin{figure}[!htbp]
    \centering
    \includegraphics[width=\columnwidth]{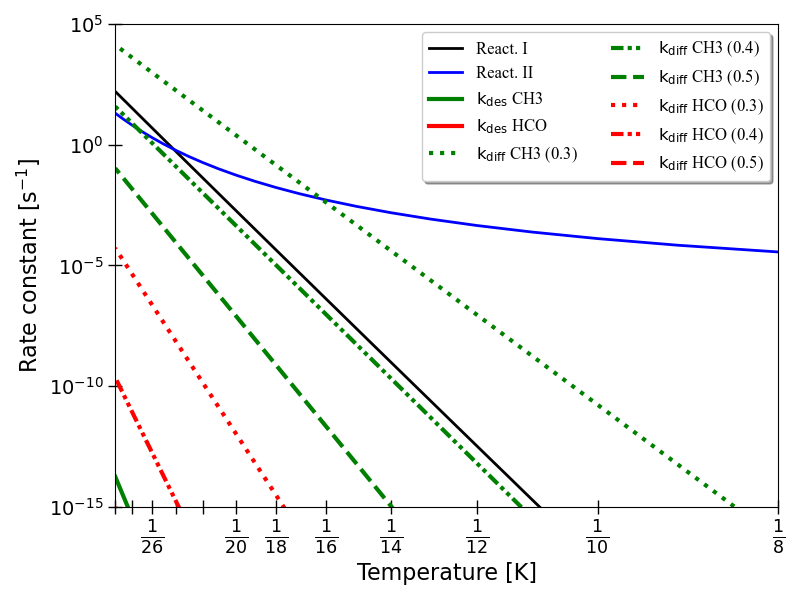}
    \caption{Comparison of reaction, diffusion and desorption rate constants involved in the CH$_3$ + HCO system. Notice that the desorption rate of HCO is not seen as it appears at very low rate constant values. Numbers in brackets indicate the diffusion-to-desorption energy barrier ratio.}
    \label{fig:supp:hoppin_rates}
\end{figure}

Figure \ref{fig:supp:hoppin_rates} compares the rates of the radical-radical reactions with the hopping and desorption rates for each radical species, using three different criteria for the diffusion barrier, namely making it 0.3, 0.4 and 0.5 times those of desorption. 

\FloatBarrier
\section{H + CO PES and data}\label{sec:appendixHCO}

\begin{figure}[!htbb]
    \centering
    \includegraphics[width=\columnwidth]{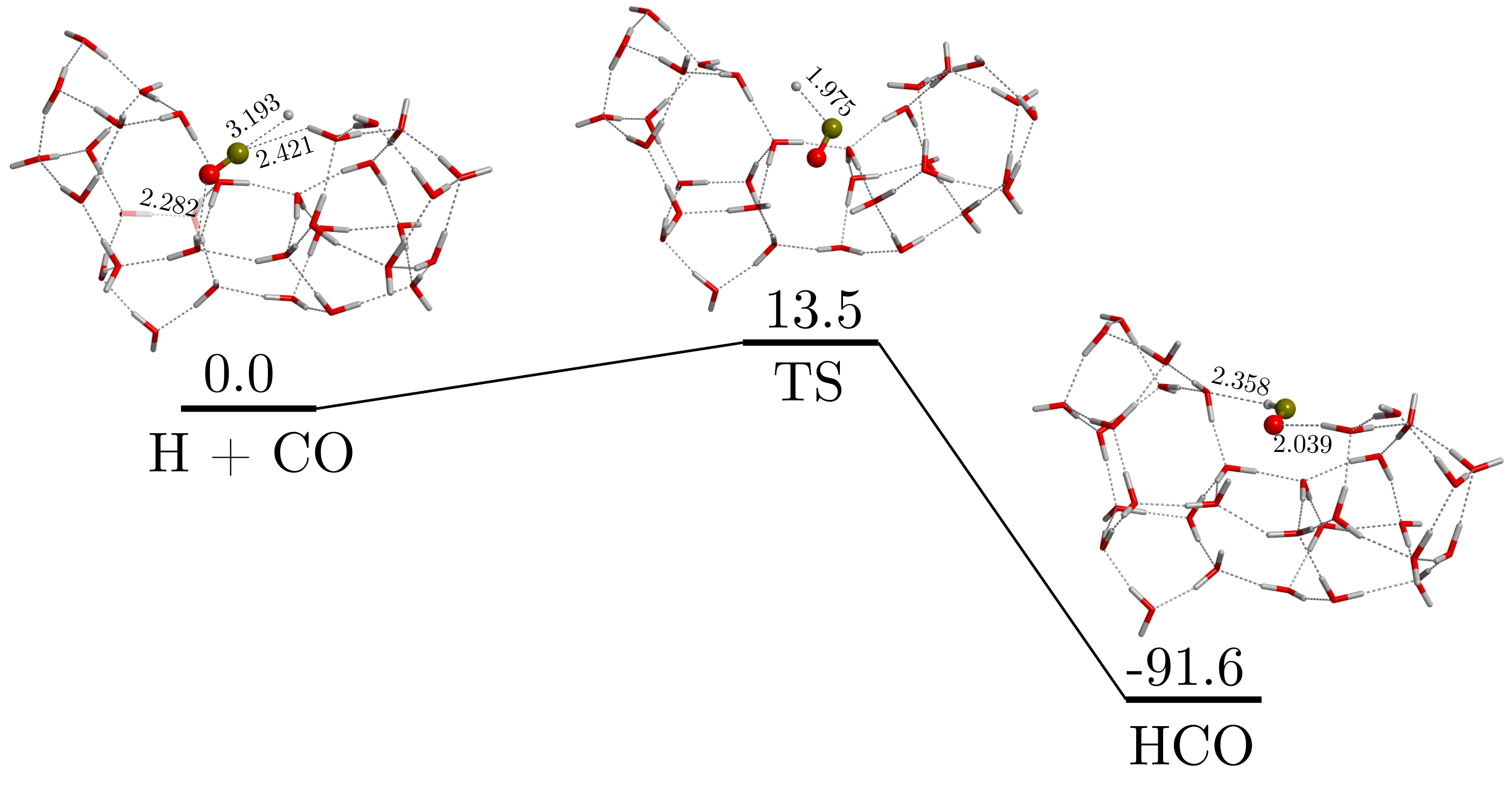}
    \caption{Potential energy surface of the H + CO $\to$ HCO reaction, in kJ mol$^{-1}$. Energies are corrected for dispersion and ZPE. Geometries and ZPE energies were obtained at UBHandHLYP-D3(BJ)/6-31+G(d,p) level and DFT energies were refined at UBHandHLYP-D3(BJ)/6-311++G(2df,2pd) level. Reactants and products were obtained by running intrinsic reaction coordinate calculations.}
    \label{fig:supp:PES_H+CO}
\end{figure}

\begin{table}[!htb]
\centering
\caption{H + CO $\to$ HCO energetic data, in Hartree, at UBHandHLYP-D3(BJ)/6-31+G(d,p) (double $\zeta$) level. $U$ is the DFT energy, $D$ is the Dispersion energy and ZPE is the zero-point energy. DFT energies were refined by performing single point calculations on double $\zeta$ geometries at UBHandHLYP-D3(BJ)/6-311++G(2df,2pd) (triple $\zeta$) level. Energy units are Hartree (1.0 Hartree are $\sim$ 2625.5 kJ mol$^{-1}$). }
\label{tab:HCO_energetics}
 \resizebox{\columnwidth}{!}{%
    \begin{tabular}{cccc}
    \hline\hline
        $H + CO$ & Reactant & TS & Product \\ \hline
        $U$ (double $\zeta$)& -2558.84099479 & -2558.83665194 & -2558.88783666 \\
        $D$ (double $\zeta$)& -0.08428582 & -0.08266012 & -0.08407096 \\
        ZPE (double $\zeta$)& 0.86306000  & 0.86239000 & 0.872578000 \\
        $U$ (triple $\zeta$)&-2559.79323262 & -2559.78741062 & -2559.83765264 \\
        \hline\hline
    \end{tabular}%
}
\end{table}

\begin{table}[!htb]
\centering
\caption{H$\cdots$CO transition state data. Units are the usual for a {\sc Gaussian16} output: frequencies in cm$^{-1}$, IR intensities in KM/Mole, reduced masses in AMU and force constants in mDyne/\r{A}. At UBHandHLYP-D3(BJ)/6-31+G(d,p) level.}
\label{tab:HCO_TS_Data}
    \begin{tabular}{ccccc}
    \hline\hline
                 & $i\nu$ & red mass & F ctn & IR int \\ \hline
        $H + CO$ & 605.7354 & 1.1042 & 0.2387 & 14.8487 \\ \hline\hline
    \end{tabular}
\end{table}

\FloatBarrier
\section{Radical-radical TS data and PES:}
\begin{table}[!htbp]
\centering
\caption{CH$_3$ + HCO energetic data, in Hartree, at UBHandHLYP-D3(BJ)/6-31+G(d,p) level (double $\zeta$). $U$ is the DFT energy, $D$ is the Dispersion energy and ZPE is the zero-point energy. DFT energies were refined by performing single point calculations on double $\zeta$ geometries at UBHandHLYP-D3(BJ)/6-311++G(2df,2pd) (triple $\zeta$) level. Energy units are Hartree (1.0 Hartree are $\sim$ 2625.5 kJ mol$^{-1}$).}
\label{tab:CH3-HCO_energetics}
 \resizebox{\columnwidth}{!}{%
    \begin{tabular}{lccc}
    \hline\hline
        React. {\rm I} & Reactant & TS & Product \\ \hline
        $U$ (double $\zeta$) & -2675.130087 & -2675.127479 & -2675.266272 \\
        $D$ (double $\zeta$) & -0.091309097 & -0.091229233 & -0.088094692 \\
        ZPE (double $\zeta$) & 0.933864     & 0.933528     & 0.94406 \\
        $U$ (triple $\zeta$) &-2676.122287 & -2676.119848 & -2676.253986\\
        \hline
        React. {\rm II} & Reactant & TS & Product \\ \hline
        $U$ (double $\zeta$)& -2675.130087 & -2675.12475 & -2675.255587  \\
        $D$ (double $\zeta$)& -0.091309097 & -0.092369778 & -0.091838262  \\
        ZPE (double $\zeta$)& 0.933864 & 0.933255 & 0.938711 \\    
        $U$ (triple $\zeta$)& -2676.122287 & -2676.118937& -2676.250317\\
        \hline\hline
    \end{tabular}%
    }
\end{table}

\begin{table}[!htbp]
\centering
\caption{Features of the transition states studied in this work. Units are the usual for a {\sc Gaussian16} output: frequencies in cm$^{-1}$, IR intensities in KM/Mole, reduced masses in AMU and force constants in mDyne/\r{A}. At UBHandHLYP-D3(BJ)/6-31+G(d,p) level. }
\label{tab:supp:dataTSs}
\resizebox{\columnwidth}{!}{%
\begin{tabular}{ccccccccc}
\hline\hline 
 & \multicolumn{4}{c|}{reaction {\rm I}} & \multicolumn{4}{c}{reaction {\rm II}} \\ \cline{2-9} 
 & $i\nu$ & red mass & F ctn & \multicolumn{1}{c|}{IR int} & $i\nu$ & red mass & F ctn & IR int \\ \hline
\multicolumn{1}{c|}{CH$_3$ + HCO} & 91.0685 & 4.7645 & 0.0233 & \multicolumn{1}{c|}{0.5366} & 168.6488 & 1.8100 & 0.0303 & 16.6868 \\ 

\hline \hline
\end{tabular}%
}
\end{table}

\begin{figure}[!htbb]
    \centering
    \includegraphics[width=\columnwidth]{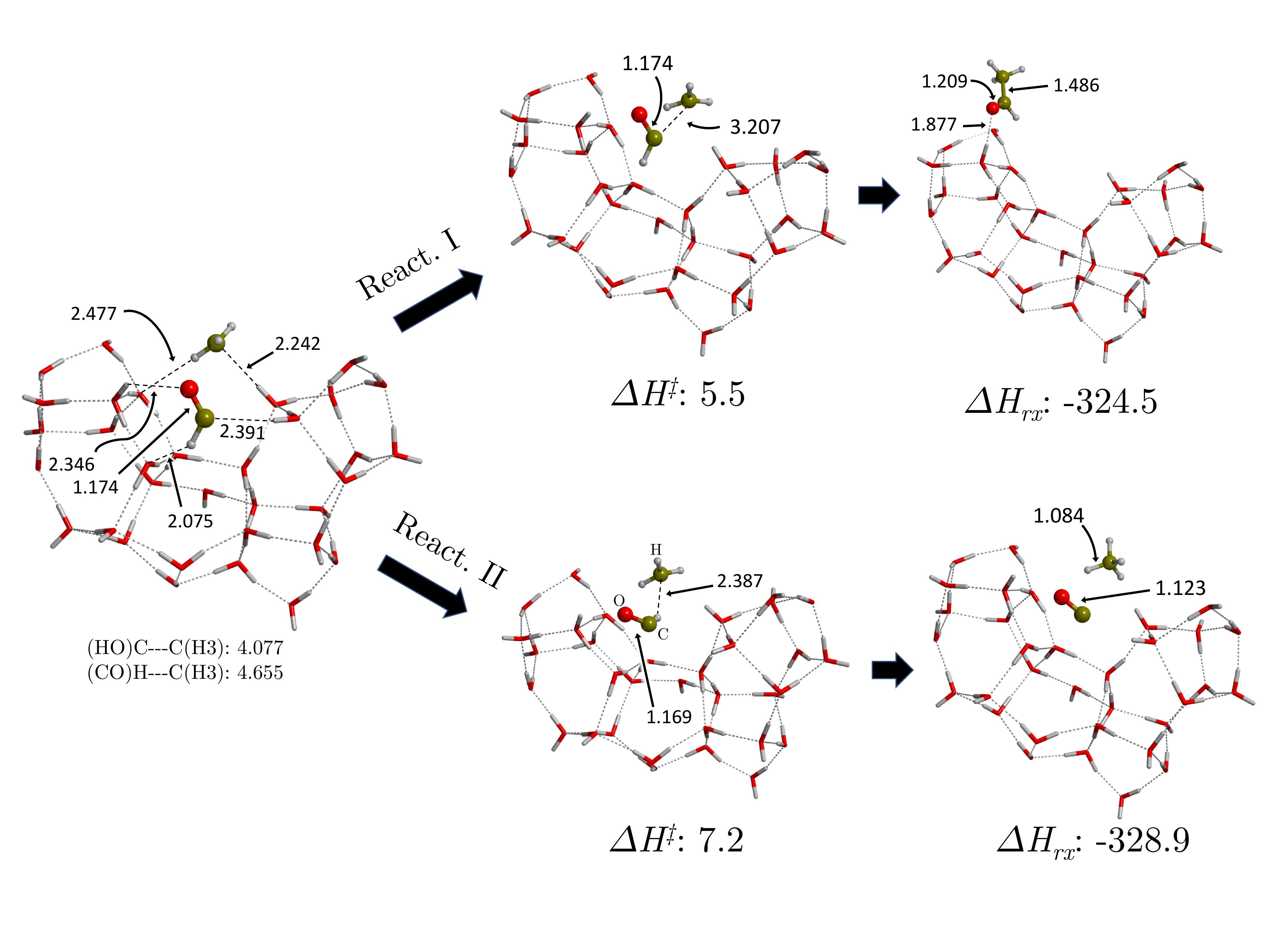}
    \caption{Potential energy surfaces Reacts. {\rm I} and {\rm II}. Geometries and ZPE energy correction was obtained at UBHandHLYP-D3(BJ)/6-31+G(d,p) level, DFT energy was refined at UBHandHLYP-D3(BJ)/6-311++G(2df,2pd) level. Energy units are in kJ mol$^{-1}$.}
    \label{fig:my_label}
\end{figure}

\FloatBarrier
\vspace*{0.3cm}
\section{Efficiency figures, separated by \texorpdfstring{$E_{diff}/E_{des}$}{E\_diff/E\_des} ratios}

\begin{figure}[!htbp]
    \centering
    \includegraphics[width=0.80\columnwidth]{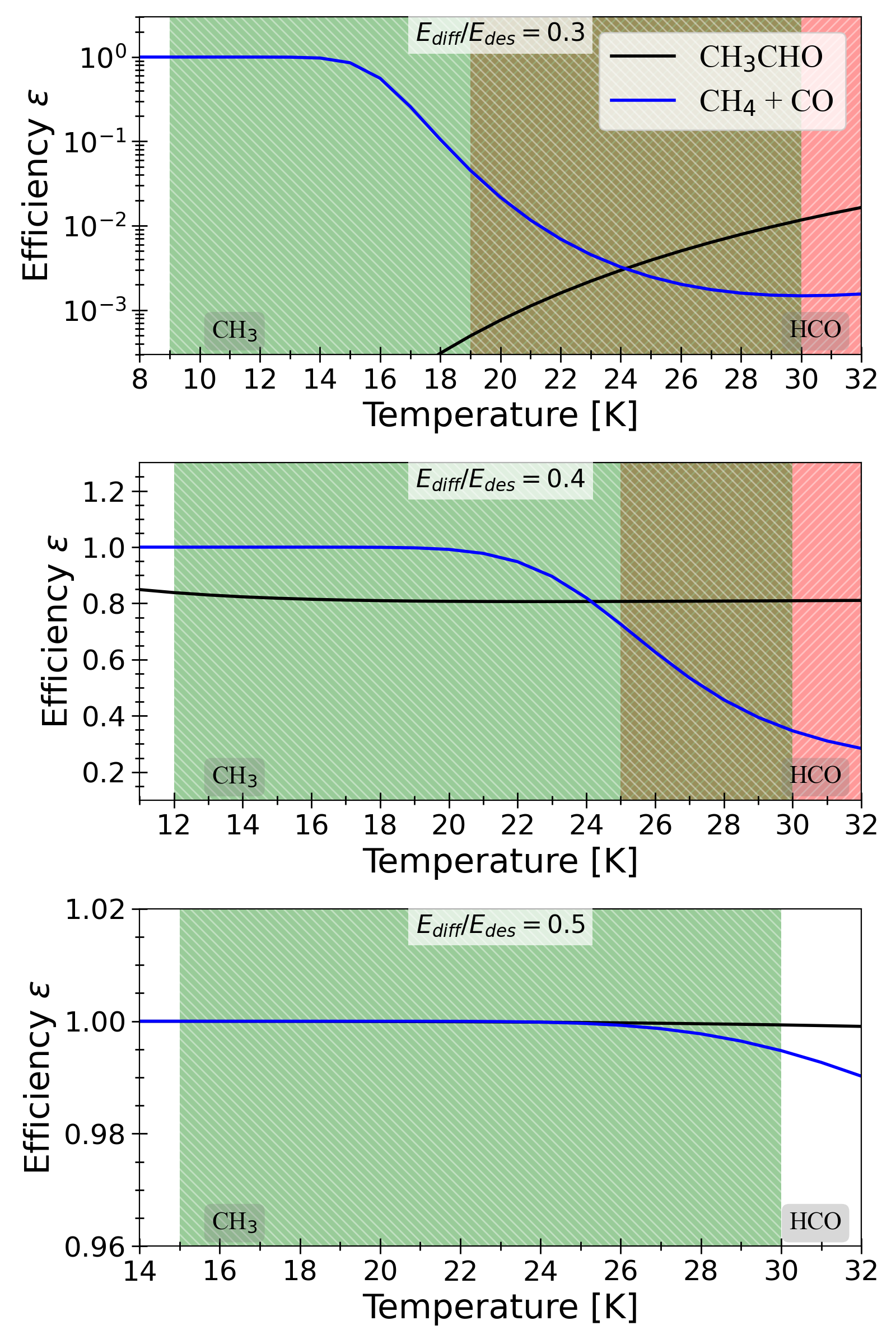}
    \caption{CH3 + HCO reaction efficiencies $\varepsilon$ (Eq. \ref{eqn:our_rx_efficiency}), assuming diffusion barriers 0.3, 0.4 and 0.5 times those of desorption (panels from top to bottom). The green-colored regions indicate the diffusion and desorption temperatures limits of CH3, while the red ones are the same for HCO.}
    \label{fig:annex:efficiency_per_ratio}
\end{figure}

\FloatBarrier
\section{Calculation of diffusion and desorption temperatures}

\begin{figure}[!htbp]
    \centering
    \includegraphics[width=\columnwidth]{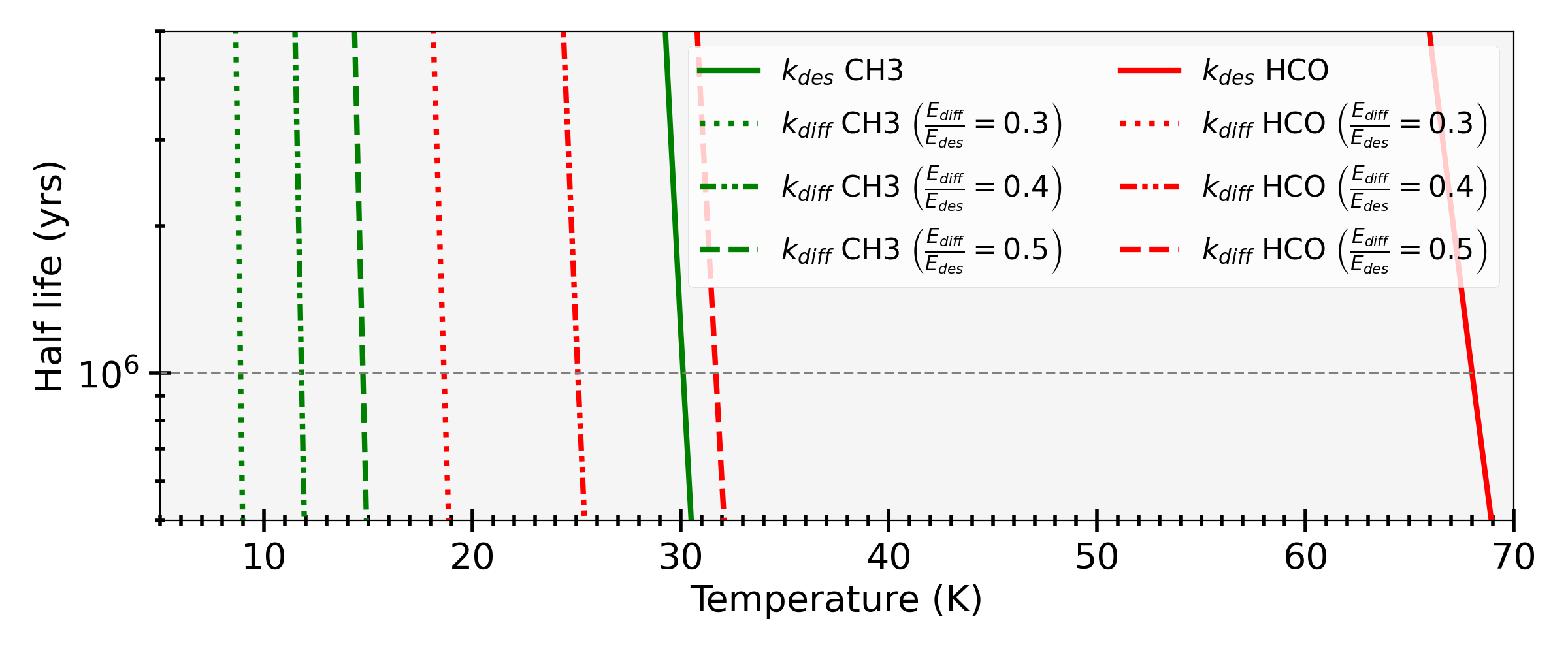}
    \caption{Diffusion and diffusion temperatures of CH$_3$ and HCO assuming a half-life of 1 Myrs for desorption.}
    \label{fig:annex:diff_des_temps}
\end{figure}

Half-lives are calculated from the rate constants ($k_i$, with $i$ being either the diffusion or desorption of radicals) following

$$
N_{1/2} = N_0 \exp(-k_i t) \qquad \Longrightarrow \qquad t_{1/2} = \frac{\ln(2)}{k_i},
$$

and shown in Fig. \ref{fig:annex:diff_des_temps}.

\FloatBarrier
\section{Fittings to \texorpdfstring{$k_{aeb}$}{k\_aeb} and \texorpdfstring{$\varepsilon$}{epsilon}:}



\begin{figure}[!htbp]
    \centering
    \includegraphics[width=0.8\columnwidth]{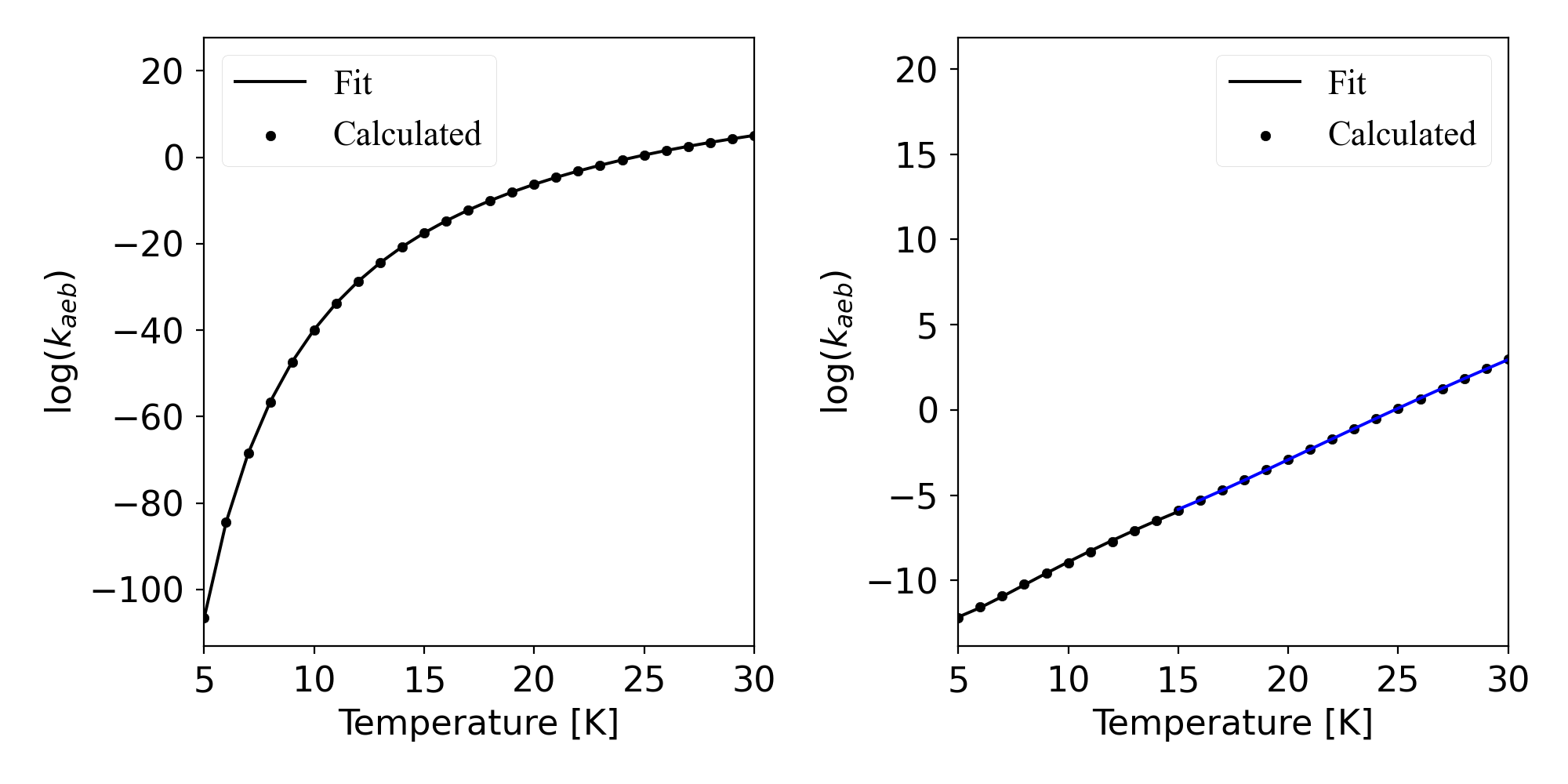}
    \caption{Fittings to the computed rate constants (Figure \ref{fig:unimoRatesNonDeut}) with Eq. \ref{eqn:coefficients} for reactions {\rm I} (left hand side panel) and {\rm II} (left hand side panel).}
    \label{fig:annex:fittings_k}
\end{figure}

\begin{figure}[!htbp]
    \centering
    \includegraphics[width=\columnwidth]{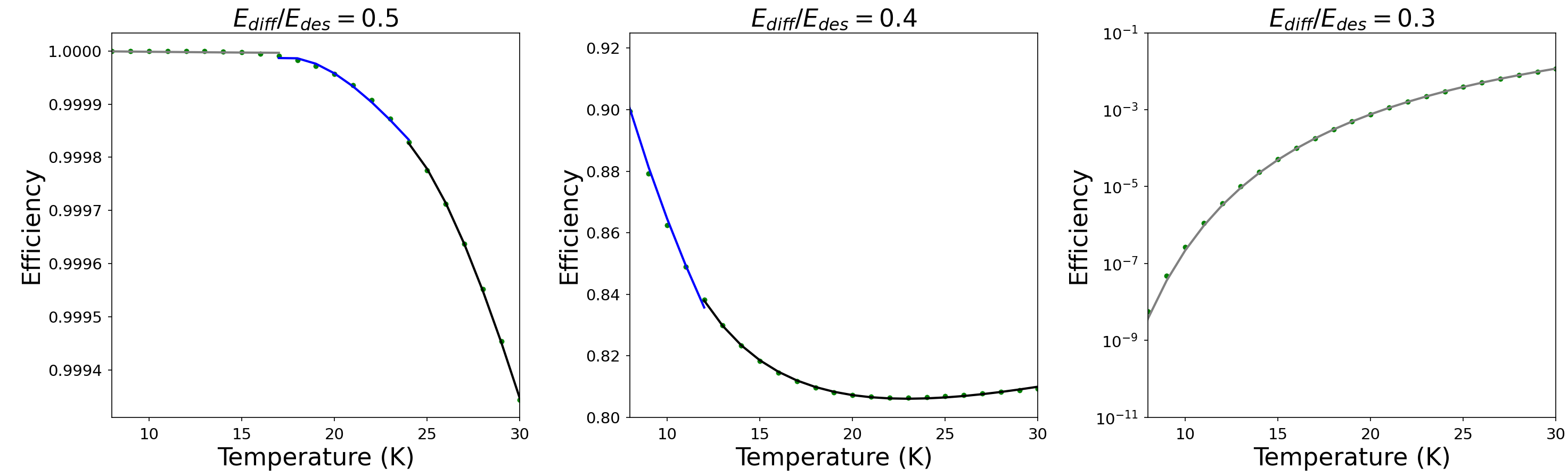}
    \caption{Fittings (solid lines) to the computed efficiency factors (points) using Eq. \ref{eqn:coefficients} for acetaldehyde using $E_{diff}/E_{des}=0.5$, $0.4$ and $0.3$ (left to right panels).}
    \label{fig:annex:fittings_Rc}
\end{figure}

\begin{figure}[!htbp]
    \centering
    \includegraphics[width=\columnwidth]{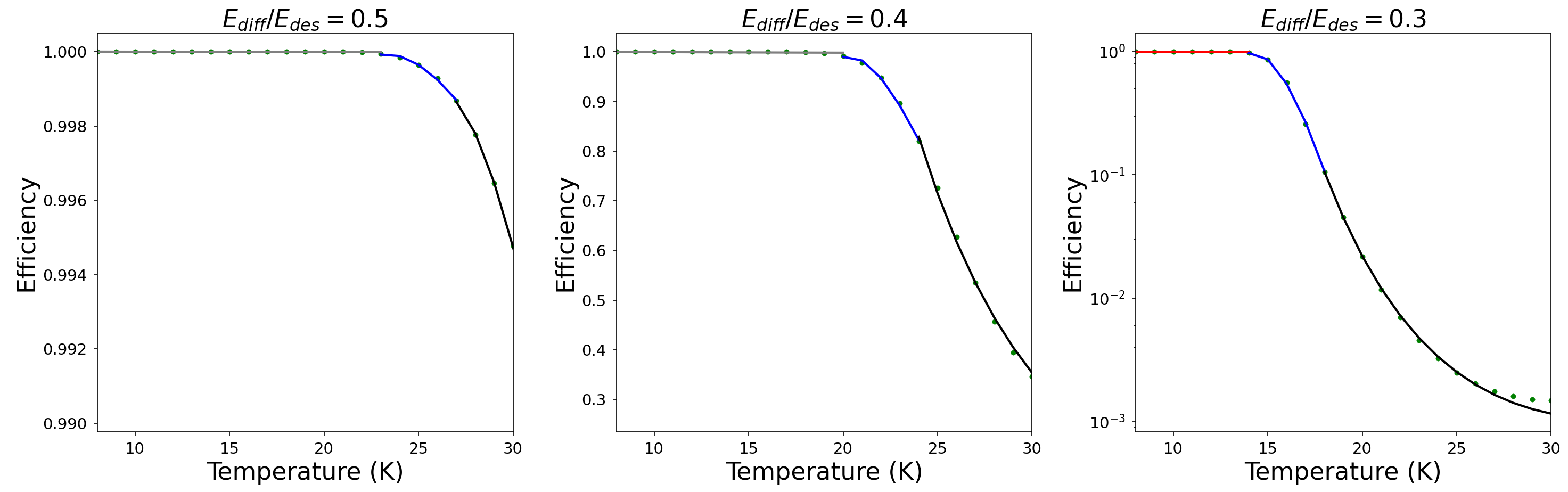}
    \caption{Fittings (solid lines) to the computed efficiency factors (points) using Eq. \ref{eqn:coefficients} for CO + CH$_4$ formation using $E_{diff}/E_{des}=0.5$, $0.4$ and $0.3$ (left to right panels).}
    \label{fig:annex:fittings_dHa}
\end{figure}


\end{appendix}
\end{document}